\documentclass[twocolumn,showpacs,preprintnumbers,amsmath,amssymb,floatfix]{revtex4}
\usepackage{amsthm}
\usepackage{graphicx}
\usepackage{dcolumn}
\usepackage{bm}
\usepackage{color}

\newcommand\HR{\rule{0em}{0pt}}
\newcommand{\BM}[1]{{\mbox{\boldmath $#1$}}}
\newcommand{\be}{\begin{equation}}
\newcommand{\ee}{\end{equation}}
\newcommand{\ba}{\begin{eqnarray}}
\newcommand{\ea}{\end{eqnarray}}
\newcommand{\ban}{\begin{eqnarray*}}
\newcommand{\ean}{\end{eqnarray*}}

\newcommand{\hu}{\hat{U}}
\newcommand{\cu}{{\cal U}}
\newcommand{\hv}{\hat{V}}
\newcommand{\bu}{\bar{U}}
\newcommand{\bv}{\bar{V}}

\newcommand{\uu}[1]{{u}_{#1}}
\newcommand{\uuu}[1]{{u}^{(2)}_{#1}}

\newcommand{\n}[1]{\label{#1}}

\newcommand{\eq}[1]{(\ref{#1})}

\newcommand{\hh}{\, ,\hspace{0.5cm}}
\newcommand{\hhh}{\, ,\hspace{0.2cm}}

\newcommand{\hn}{\hspace{-0.09cm}}
\newcommand{\hnH}{\hspace{-0.045cm}H}

\begin{document}
 
\title{Interior of a Charged Distorted Black Hole}
\author{Shohreh Abdolrahimi}
\email{sabdolra@phys.ualberta.ca}
\author{Valeri P. Frolov}
\email{frolov@phys.ualberta.ca}
\author{Andrey A. Shoom}
\email{ashoom@phys.ualberta.ca}
\affiliation{Theoretical Physics Institute, University of Alberta, 
Edmonton, AB, Canada,  T6G 2G7}
\date{\today}

\begin{abstract}

We study interior of a charged, non-rotating distorted black hole. We consider 
static and axisymmetric black holes, and  focus on a special case when
an electrically charged distorted solution is obtained by the 
Harrison-Ernst transformation from an uncharged one. We demonstrate
that the Cauchy horizon of such black hole remains regular, provided
the distortion is regular at the event horizon. The shape and the
inner geometry of both the outer and inner (Cauchy) horizons
are studied. We demonstrate that there exists a duality between
the properties of the horizons. Proper time of a free fall of a test particle moving
in the interior of the distorted black hole along the symmetry axis
is calculated. We also study the property of the curvature in the
inner domain between the horizons. Simple relations between the 4D 
curvature invariants and the Gaussian curvature of the outer and inner horizon surfaces are found.

\end{abstract}

\pacs{04.20.Dw, 04.20.Cv, 04.70.Bw \hfill  
{\bf Alberta-Thy-07-09}}
\maketitle
\section{\label{sec:level1}INTRODUCTION}

In this paper we study how the distortion of a charged, static black hole
generated by axisymmetric, static matter distribution in its exterior
region affects its interior. This paper is a direct generalization of
a similar study for the distorted neutral black hole interior
performed in \cite{FS}.

Structure and properties of the charged and/or rotating black hole
interior is a subject that has attracted a lot of interest during
past 30 years (see e.g. \cite{FN} and references therein). Analytic
continuation of the Reissner-Nordstr\"om (RN) and Kerr solutions results
in the existence of infinitely many new `universes' in the black
holes interior. However, the region containing these new `universes'
lies in the future of the Cauchy horizon, a null hypersurface beyond
which predictability breaks down. A natural question is
whether these `universes' are accessible to an observer traveling in
the interior of the black hole. That is why the issue of the Cauchy
horizon stability is so important. Observers traveling along a
timelike world line receive an infinitely blue-shifted radiation
when they approach the horizon. Penrose \cite{Pen} used these facts
to argue that small perturbations produced in the black hole exterior
grow infinitely near the Cauchy horizon. The evolution of small
perturbations inside charged black holes  was analyzed  in
\cite{Per1}, \cite{Per2}, \cite{Per3}. These results confirm 
Penrose's intuitive arguments.

If one considers ingoing radiation only and neglects backscattered
radiation, then the resulting Cauchy horizon singularity is weak. 
Namely, the Kretschmann invariant calculated on the Cauchy horizon is finite. 
A freely falling observer detects an infinite increase of energy density, but tidal forces remain finite
as the observer crosses the Cauchy horizon \cite{Eli}, \cite{Kin}.
Such singularity is called the whimper singularity.
However, in a realistic situation, when both incoming and outgoing
radiation are present, the curvature grows infinitely near the Cauchy
horizon. This was demonstrated by Poisson and Israel \cite{IP} who
considered the outgoing and ingoing radiation simulated as two
non-interacting radial streams of ingoing and outgoing lightlike
particles following null geodesics. Poisson and Israel showed that
such radiation results in an infinite growth of the black hole internal
mass parameter and divergence of the Weyl scalar. They
called this effect the {\em mass inflation}. Mass inflation for a slowly rotating, 
charged black hole was discussed in \cite{Ham}. Later, Ori constructed
an exact, simplified solution describing this effect \cite{Ori}. Using
his solution Ori showed that the {\em mass inflation} singularity is
weak enough. Namely, the tidal forces calculated at the Cauchy
horizon diverge in the reference frame of a freely falling observer,
but their integral along the world line of the observer remains
finite. It means that freely falling observers might in fact cross
the Cauchy horizon. For more detailed discussion see e.g. \cite{B1}-\cite{Ham2}.
Early numerical analysis of the Cauchy horizon 
stability predicted its destruction as a result of classical
instability \cite{Gne}. Later, analytical \cite{Bon}, \cite{Fla}, and
numerical \cite{Bra} discussions did not confirm this result. The
mass inflation phenomenon may shed light on the Cauchy horizon
stability problem. However, further investigation is necessary.

Although rotating black holes are of real astrophysical interest,
charged black holes are often considered in the publications. The
reason for this is simple: a charged black hole also has Cauchy
horizon, but its spherical geometry makes an analysis easier.
However, even in this case such model is very simplified,
for in the realistic world there always exists some
matter outside the black hole. This matter distorts the gravitational
field of the black hole. What is important, that this distortion
generated by the matter distribution in the exterior of the black
hole occurs not only outside the black hole, but also affects its
interior. Since the region near the Cauchy horizon is `fragile' and
`vulnerable', it is interesting to analyze how such external matter
affects the properties of the black hole Cauchy horizon. This is one of the
questions we address in our paper. We shall make several assumptions
simplifying the analysis. Namely, we assume that the distortions of
the black hole are static and axisymmetric. Moreover, we consider a
special class of charged distorted black hole solutions which can be
generated by the Harrison-Ernst transformation
\cite{Har}, \cite{Ernst} from a neutral distorted black hole
metric. This class includes a large variety of solutions which can be
presented in an explicit form. 

We always assume that  in the vicinity of the black hole and in its
interior the Einstein-Maxwell equations are satisfied, and the matter
disturbing the black hole is located in the black hole exterior. The
matter sources are described by the corresponding energy-momentum
tensor which has to be included in the Einstein-Maxwell equations. 
To avoid this one can
`move' these sources to infinity. The `price' for this is that the
corresponding spacetime is not  asymptotically flat
anymore. In our description of a distorted black hole we follow
\cite{Geroch} and adopt that approach. 

Our main problem is to study how the black hole interior is distorted by
the external fields. In particular, we shall study distortion of the
inner (Cauchy) horizon and its relation to the distortion of the
outer (event) horizon. Let us emphasize that our consideration is completely
classical, and we do not consider quantum effects which may play an
important role in the charged black hole interior. Discussion of
these effects can be found e.g. in \cite{Gurt}-\cite{AKT2}.

It should be emphasized that the study of the black hole interior is
a dynamical problem. The geometry of the black hole interior is
similar to the geometry of a contracting, anisotropic, homogeneous
`universe'. To study how the evolution of this `universe' is modified
by an external influence, one must study first the modification of
the external geometry of the black hole and use these results to find
the corresponding modification of the geometry of the event horizon.
This gives the initial data which determine the evolution of the
black hole interior. In this paper we study a simple case when the
distortion of the black hole in the exterior region is both
stationary and axisymmetric. A similar problem for the neutral black
hole was studied earlier in \cite{FS}. 

This paper is organized as follows. Section II collects the results
concerning the charged distorted black hole solution generated by
the Harrison-Ernst transformation technique. We remind these results
mainly in order to fix the notations we use in the main part of the
paper. In Section III we establish special duality relations between properties of the
inner and outer horizons for the charged distorted black hole.
In Sections IV and V we study the Gaussian curvature of the horizon
surfaces  and present their isometric embedding diagrams. In Section
VI we discuss how the black hole distortion affects the maximal proper time of a free fall
of a test particle moving along the axis of symmetry in the black
hole interior. In Section VII we establish a relation between
the spacetime curvature invariants near the horizons and their
Gaussian 2D curvatures.  We summarize and discuss
our results in Section XIII. Necessary details are included in the
appendix. In this paper we use the units where $G=c=1$, and the
sign conventions adopted in \cite{MTW}.

\section{Metric of a Distorted RN Black Hole}

\subsection{Static, axisymmetric Einstein-Maxwell space-time}

In this Section following \cite{Bre1}, \cite{Bre2}, \cite{Step} we present
a solution for static, axisymmetric distorted charged black hole. This
solution is obtained by applying the Harrison-Ernst transformation  \cite{Har}, \cite{Ernst} to
the Weyl metric of a distorted vacuum black hole. Here we reproduce
the basic relations, mainly in order to explain notations we shall
use later.

The metric of charged distorted black hole is a special solution of
the Einstein-Maxwell equations
\ba\n{3}
R_{\alpha\beta}&=&8\pi\, T_{\alpha\beta},\\
\nabla_{\beta}F^{\alpha\beta}&=&0\hhh \nabla_{[\alpha}F_{\beta\gamma]}=0,\\
8\pi\, T_{\alpha\beta}&=&2F_{\alpha}^{\,\,\gamma}F_{\beta\gamma}-\frac{1}{2}g_{\alpha\beta}
F_{\gamma\delta}F^{\gamma\delta}.
\ea
Here, $F_{\alpha\beta}=\nabla_{\alpha}A_{\beta}-\nabla_{\beta}A_{\alpha}$,
and $A_{\alpha}$ is the electromagnetic 4-potential. 
The nabla stands for covariant derivative defined with respect to the metric $g_{\alpha\beta}$. 

Before we proceed with
description of a charged distorted black hole, let us make a few
remarks about charged black hole solution in the absence of
distortions. This is the well-known Reissner-Nordstr\"om solution
(see e.g. \cite{Chandrabook})
\ba\n{RN1a}
ds^2&=&-Fdt^2+F^{-1}dr^2+r^2(d\theta^2+\sin^2\theta d\phi^2),\\ 
F&=&1-{2M\over r}+{Q^2\over r^2}\hhh A_{\alpha}=-\Phi_0 \delta_{\alpha}^{\,\,\,\, t}\hhh\Phi_0=\frac{Q}{r}.
\ea  
Here, $M$ is the black hole mass, and $Q$ is its electric charge.
We shall consider non-extremal black holes with
$|Q|<M$. The spacetime is static and asymptotically flat.  It has a
timelike singularity at $r=0$. The black hole horizons are defined
by $r_{\pm}=M\pm\sqrt{M^2-Q^2}$, where the upper sign stands for the
event horizon, and the lower sign stands for the Cauchy horizon. Correspondingly, we denote these horizons as ${\cal H}^{(\pm)}$.

It is convenient to make the following coordinate transformation
\ba\n{re}
r=M(1+p\eta)\hhh p=\frac{\sqrt{M^2-Q^2}}{M}\hhh
\eta\in(-1/p, \infty),
\ea
and to rewrite the Reissner-Nordstr\"{o}m solution in the following form
\ba\n{RN2b}
ds^2&=&-\frac{p^2(\eta^2-1)}{(1+p\eta)^2}dt^2+M^2(1+p\eta)^2\nonumber\\
&\times&\biggl[\frac{d\eta^2}{\eta^2-1}+d\theta^2+\sin^2\theta d\phi^2\biggl],\\
\Phi_0&=&{\sqrt{1-p^2}\over {(1+p\eta)}}.\n{Phi}
\ea
In these new coordinates $\eta=\eta_{\pm}=\pm1$ corresponds to the horizons
of metric \eq{RN2b}, and $\eta=-1/p$ corresponds to the black hole singularity.

The general form of static, axisymmetric metric in prolate spheroidal
coordinates $(\eta, \cos\theta)$ reads
\ba\n{1}
ds^2&=&-e^{2U}dt^2+
M^2p^2e^{-2U}\biggl[e^{2V}(\eta^2-\cos^2\theta)\biggr.\nonumber\\
&\times&\left.\left(\frac{d\eta^2}{\eta^2-1}+d\theta^2\right)
+(\eta^2-1)\sin^2\theta d\phi^2\right],
\ea 
where the metric functions $U$ and $V$ depend on $(\eta, \theta)$
coordinates. The corresponding electrostatic 4-potential is 
\be\n{2}
A_{\alpha}=-\Phi(\eta,\theta)\delta_{\alpha}^{\,\,\,\, t}.
\ee 

\subsection{The Harrison-Ernst transformation}

The Einstein-Maxwell equations for $U$ and $\Phi$ are the Ernst
equations \cite{Ernst}, which in our case of static spacetime \eq{1}
take the following form  
\ba\n{6}
\nabla \left(e^{-2U}\nabla\mathcal{E}\right)=0\hh
\nabla \left(e^{-2U}\nabla{\Phi}\right)=0.
\ea
Here, $\mathcal{E}=e^{2U}-\Phi^2$ is the Ernst potential, and ${\nabla}$ 
is the nabla operator defined with respect to the 3D flat metric
\ba\n{7}
dl^2&=&(\eta^2-\cos^2\theta)\left[\frac{d\eta^2}{\eta^2-1}
+d\theta^2\right]+(\eta^2-1)\sin^2\theta d\phi^2.\nonumber\\
\ea 
There exists a special class of solutions where the Ernst potential
$\mathcal{E}$  is an analytic function of $\Phi$. Under this
assumption equations \eq{6} imply 
\be\n{6a}
\frac{d^2\mathcal{E}}{d\Phi^2}=0.
\ee
If spacetime is asymptotically flat, we choose $U=\Phi=0$
at infinity. In this case a general solution of \eq{6a} can be written
as 
\be\n{6b}
\mathcal{E}=1-\frac{2}{\sqrt{1-p^2}}\Phi.
\ee
We shall keep this relation in our consideration. Following \cite{Ernst} it is convenient to parametrize $\mathcal{E}$ and $\Phi$ as follows
\be\n{6c}
\mathcal{E}=\frac{\xi-1}{\xi+1}\hh
\Phi=\frac{\sqrt{1-p^2}}{\xi+1},
\ee
where $\xi$ is the auxiliary Ernst potential.
Using \eq{6} one obtains the following equation for $\xi$
\be\n{8a}
(\xi^2-p^2)\nabla^2\xi-2\xi\nabla\xi\cdot\nabla\xi=0.
\ee

In the absence of electric field, $\Phi=0$, the Ernst equation \eq{6} is
\be\n{8b}
\bar{\mathcal{E}}\nabla^2\bar{\mathcal{E}}=
\nabla\bar{\mathcal{E}}\cdot\nabla\bar{\mathcal{E}},
\ee
where $\bar{\mathcal{E}}=e^{2\bu}$, and $\bu$ corresponds to
vacuum uncharged solution. In this case one can also use parametrization \eq{6c} which gives
\be\n{9}
\bar{\mathcal{E}}=\frac{\bar{\xi}-1}{\bar{\xi}+1},
\ee
and the Ernst equation \eq{8b} takes the form
\be\n{9a}
(\bar{\xi}^2-1)\nabla^2\bar{\xi}-2\bar{\xi}\nabla\bar{\xi}\cdot\nabla\bar{\xi}=0.
\ee
Comparing \eq{8a} and \eq{9a} we can derive the relation between the vacuum and the electrostatic Ernst potentials. This is the Harrison-Ernst transformation:
\be\n{9b}
\xi=p\bar{\xi}.
\ee
Thus, if we know a solution to vacuum Einstein equations $\bu$, we
can apply \eq{9b} and \eq{6c} to obtain the corresponding solution $U$,
and the electrostatic potential $\Phi$ obeying the Einstein-Maxwell equations.
Namely, using expressions \eq{9b}, \eq{9} and \eq{6c} we derive 
\ba\n{rel1}
e^{2U}&=&\frac{4p^2e^{2\bu}}{[1+p-(1-p)e^{2\bu}]^2}\hhh \Phi=\frac{\sqrt{1-p^2}(1-e^{2\bu})}{1+p-(1-p)e^{2\bu}}.\nonumber\\
\ea
These expressions  determine the charged version of electrically
neutral, vacuum static solution. For example, starting with the
Schwarzschild black hole solution we can derive the Reissner-Nordstr\"om black hole. If Schwarzschild black hole is distorted by neutral exterior matter, these expressions electrically charge both, the black hole and the matter.  

In the next subsection we apply this `charging' procedure to  the
Weyl static metric describing a vacuum, axisymmetric distorted black
hole, and obtain electrically charged distorted black hole.
We discuss the corresponding metric in the next subsection.

\subsection{Charged distorted black hole}

Now we are ready to present a solution for a charged, axisymmetric
distorted black hole. Following to the procedure presented in the
previous subsection we start with the vacuum solution representing
axisymmetric distorted Schwarzschild black hole which we write in the
form \cite{FS}, \cite{TD}
\ba\n{9c}
ds^2&=&-e^{2\bu}dt^2+M^2e^{-2\bu}
\biggl(e^{2\bv}(\eta^2-\cos^2\theta)\biggr.\nonumber\\
&\times&\left.\left[\frac{d\eta^2}{\eta^2-1}+d\theta^2\right]
+(\eta^2-1)\sin^2\theta d\phi^2\right).\\
e^{2\bu}&=&\frac{\eta-1}{\eta+1}e^{2\hu}\hhh
e^{2\bv}=\frac{\eta^2-1}{\eta^2-\cos^2\theta}e^{2\hv}.\n{v} 
\ea 
For undistorted Schwarzschild solution $\hu=\hv=0$. For the distorted
metric the
vacuum Einstein equations for $\hu$ and $\hv$ distortion fields imply
\be\n{10}
(\eta^2-1)\hu_{,\eta\eta}+2\eta\hu_{,\eta}+\hu_{,\theta\theta}+\cot\theta \hu_{,\theta}=0,
\ee
\ba
\hv_{,\eta}&=&N\left(\eta[(\eta^2-1)\hu_{,\eta}^2-\hu_{,\theta}^2]+2(\eta^2-1)\cot\theta\hu_{,\eta}\hu_{,\theta}\right.\nonumber\\
&+&\left.2\eta\hu_{,\eta}+2\cot\theta\hu_{,\theta}\right),\n{11a}\\
\hv_{,\theta}&=&-N\left((\eta^2-1)\cot\theta[(\eta^2-1)\hu_{,\eta}^2-\hu_{,\theta}^2]\right.\nonumber\\
&-&\left.2\eta(\eta^2-1)\hu_{,\eta}\hu_{,\theta}+2(\eta^2-1)\hu_{,\eta}-2\eta\hu_{,\theta}\right)\n{11b}.
\ea 
Here, $N=\sin^2\theta(\eta^2-\cos^2\theta)^{-1}$, and comma
stands for a partial derivative. Once the solution to equation \eq{10} is found, $\hv$ can be determined by integration of \eq{11a}, \eq{11b}. Details of derivation of $\hu$ can be found for example in \cite{Chandrabook}, \cite{FS}. Regularity of the distorted black hole horizon implies that $\hu$ can be decomposed over the Legendre polynomials of the first kind
\ba
\hu&=&\sum_{n\geqslant0}a_nP_n(\eta)P_n(\cos\theta).
\ea 
Thus, $\hu$ and its derivatives are everywhere regular. Using this decomposition one can write the distortion field in equivalent form \cite{Bre1}, \cite{Bre2}
\ba
\hu&=&\sum_{n\geqslant0}c_nR^nP_n,\n{2.6a}\\
P_n&=&P_n\left(\eta\cos\theta/R\right)\hhh
R=(\eta^2-\sin^2\theta)^{1/2}\, .\n{PR}
\ea
Here, the constant coefficients $c_n$'s define the distortion field.
We call these coefficients the {\em multipole
moments} \cite{f1}. The multipole
moments uniquely characterize the distortion. 
Later we discuss some examples illustrating nature of 
distortion defined by the lowest multipole moments.

The distortion field $\hv$ can be written in a closed form as a sum of two
terms $\hv=\hv_1+\hv_2$ (see e.g. \cite{Bre1,Bre2}). The first term, $\hv_1$, is linear, and the second one, $\hv_2$, is quadratic in $c_n$'s
\ba
\hv_1&=&\sum_{n\geqslant1}c_n\sum_{l=0}^{n-1}\left[\cos\theta-\eta-(-1)^{n-l}(\eta+\cos\theta)\right]R^lP_l,\nonumber\\
\n{v1}\\
\hv_2&=&\sum_{n,k\geqslant1}\frac{nkc_nc_k}{n+k}R^{n+k}(P_nP_k-P_{n-1}P_{k-1}).\n{v2}
\ea
An equilibrium of the black hole with respect to the distortion
fields means that the distortion field $\hu$ takes the same values at
the points of the symmetry axis on the black hole outer horizon (see e.g. \cite{Geroch}),
\be
\hu(\eta=1,\theta=0)=\hu(\eta=1,\theta=\pi)\equiv u_0.
\ee
We can rewrite this condition in terms of the multipole moments. Using \eq{2.6a}, \eq{PR} and the property of the Legendre polynomials, 
\be\n{Lp}
P_n(\pm1)=(\pm1)^n,
\ee
the equilibrium condition reads
\be
\sum_{n\geqslant0}c_{2n+1}=0,
\ee
and one has
\be\n{sumc}
u_0=\sum_{n\geqslant0}c_n=\sum_{n\geqslant0}c_{2n}\, .
\ee
Thus, a static, axisymmetric, distorted black hole is at equilibrium if the sum of odd multipole moments of the distortion vanishes. The equilibrium condition implies the local flatness (absence of conical singularities) along the  symmetry axis of the black hole. Namely,
\be\n{lf}
\hv(\eta,\theta=0)=\hv(\eta,\theta=\pi)=0.
\ee

To obtain a charged version of the distorted black hole it is
sufficient to derive $U$ and $\Phi$ from $\bar{U}$ (see \eq{v}, \eq{2.6a}) using the
Harrison-Ernst transformation \eq{rel1}. We have: 
\ba\n{RNU}
e^{2U}&=&\frac{4p^2(\eta^2-1)e^{2\hu}}{[(1+p)(\eta+1)-(1-p)(\eta-1)e^{2\hu}]^2},
\ea
\ba\n{RNF}
\Phi&=&\frac{\sqrt{1-p^2}[\eta+1-(\eta-1)e^{2\hu}]}{(1+p)(\eta+1)-(1-p)(\eta-1)e^{2\hu}}.
\ea

Remarkably, the Harrison-Ernst transformation does not alter
equations \eq{11a} and \eq{11b}. 
Thus, $U$ and $\Phi$, given by \eq{RNU}, \eq{RNF}, and $\bv$, which is determined by 
\eq{v}, \eq{v1} and \eq{v2}, 
solve the corresponding Einstein-Maxwell equations. The axisymmetric distorted
RN solution is given by \eq{1}, \eq{2} with \eq{RNU}, \eq{RNF} and
$V=\bv$. A more general case of distorted, electrically
charged, rotating black hole  is considered in \cite{Bre2}.

\subsection{Dimensionless form of the metric}

The black hole metric contains only one essential dimensional parameter, say its mass, while all other parameters can be presented in dimensionless form. It is convenient to write metric \eq{1} in the following dimensionless form adopted to the black hole horizons ${\cal H}^{(\pm)}$:
\ba
ds^2&=&\Omega^2_{\pm}dS^2_{\pm},\n{2.1}\\
dS_{\pm}^2&=&-\frac{\eta^2-1}{\Delta_{\pm}}e^{2\cu}dT_{\pm}^2+\frac{\Delta_{\pm}}{\eta^2-1}e^{-2\cu+2\hv}d\eta^2\nonumber\\
&+&\Delta_{\pm} e^{-2\cu}\left(e^{2\hv}d\theta^2+\sin^2\theta d\phi^2\right)\, ,\n{dmet}\\
\Omega_{\pm}&=&M(1\pm p)e^{\mp u_0}=M'(1\pm p').
\ea
 
For $dS_{\pm}^2$, $T_{\pm}=\kappa_{\pm} t$, and $\kappa_{\pm}$ is the surface gravity  
\be\n{2.4}
\kappa_{\pm}=\frac{(1+p')e^{u_0}-(1-p')e^{-u_0}}{2M'(1\pm p')^2}\, .
\ee
We also use the following expressions for the metric functions $\Delta_{\pm}$ and $\cu$
\ba 
\Delta_{\pm}&=&\frac{\delta^{\pm1}}{4\delta}\left[\eta+1-\delta e^{2\cu} (\eta-1)\right]^2\,,\\
\cu&=&\hu -u_0\hhh \delta=\delta_0e^{2u_0}=\frac{1-p}{1+p}e^{2u_0}=\frac{1-p'}{1+p'}\,.\nonumber\\
\n{dfu}
\ea
Together with the original parameters $M$ and $p$ it is convenient to use the related parameters
\ba
M'&=&\frac{M}{2}\left[(1+p)e^{-u_0}+(1-p)e^{u_0}\right]\,,\\
p'&=&\frac{\sqrt{M'^2-Q^2}}{M'}\,.
\ea
In the absence of distortion $M'=M$ is the Komar mass of RN black hole measured at asymptotic infinity. In the case $Q=0$, $M'$ is the local mass of a distorted Schwarzschild black hole defined in \cite{Geroch}.

The coordinate $\eta$ changes from $\eta=\infty$ (a spatial infinity) to the region of $\eta<-1$ where the spacetime singularity is located (see subsection E).
As in the case of RN black hole \eq{RN2b}, the horizons of metric \eq{dmet} are defined by $\eta=\eta_{\pm}=\pm1$. As we mentioned earlier, we shall use the notation ${\cal H}^{(\pm)}$ for the outer ($+$), and for the inner ($-$) horizons. To indicate that a dimensional quantity $(\ldots)$ is calculated at the black hole horizons ${\cal H}^{(\pm)}$, we shall use a superscript $(\pm)$, and denote this quantity as $(\ldots)^{(\pm)}$ \cite{f2}.

As we shall see in the next section, the form of metric \eq{2.1} is convenient for the analysis and comparison of the properties of the inner and outer black hole horizons. 
2D metrics on the horizon surfaces can be obtained by taking $T=const.$, and $\eta=\eta_{\pm}=\pm1$ in the metric. In the next Section we show that the surface area of the outer (event) horizon calculated for the dimensionless metric $dS_+^2$ is equal to $4\pi$. Similarly, the surface area of the inner (Cauchy) horizon calculated for the metric $dS_-^2$ is also equal $4\pi$. These normalization conditions specify the form of the conformal factor $\Omega_{\pm}$ in \eq{2.1}. The `real' (dimensional) areas of the horizon surfaces are
\be\n{sur}
{\cal A}^{(\pm)}=4\pi \Omega_{\pm}^2\, ,
\ee
and the ratio of these areas is
\be
{\cal A}^{(+)}/{\cal A}^{(-)}=(\Omega_{+}/\Omega_{-})^2=\left({1+p'\over 1-p'}\right)^2\equiv \delta^{-2}\, .
\ee
In what follows, we shall discuss different geometrical objects, such as the Kretschmann invariant ${\cal K}$, the Weyl scalar ${\cal C}^2$, 
\be
{\cal K}=R_{\alpha\beta\gamma\delta}R^{\alpha\beta\gamma\delta}\hh
{\cal C}^2=C_{\alpha\beta\gamma\delta}C^{\alpha\beta\gamma\delta}\, ,
\ee
and the Gaussian curvature of the 2D horizon surface $K$.  We shall use the same notations with an index $\pm$ for an object calculated for the metric $dS_{\pm}^2$. One has
\be
{\cal K}=\Omega_{\pm}^{-4}{\cal K}_{\pm}\hhh
{\cal C}^2=\Omega_{\pm}^{-4}{\cal C}^2_{\pm}\hhh
K=\Omega_{\pm}^{-2}K_{\pm}\, .
\ee
To study the interior region we can use any of these two forms of the dimensionless metric $dS_{\pm}^2$. Certainly, the `physical' result, calculated for the metric $ds^2$ will be the same. 

The dimensionless electrostatic potential for metric \eq{dmet} is given by
\ba\n{2.8}
\Phi_{\pm}=\frac{\sqrt{\delta}\Delta^{-1/2}_{\pm}}{(e^{2u_0}-\delta)}\left[\eta+1-(\eta-1)e^{2\cu+2u_0}\right].
\ea
It is related to electrostatic potential \eq{RNF} as follows
\be\n{dimf}
\Phi=\Omega_{\pm}\kappa_{\pm}\,\Phi_{\pm}.
\ee
The non-vanishing dimensionless components of the electromagnetic field $F_{\mu \nu}$ are defined by
\ba\n{2.9}
F_{\pm T_{\pm} \eta}&=&\Phi_{\pm,\eta}=\frac{\delta^{\pm1/2}}{\Delta_{\pm}}e^{2\cu}[(1-\eta^2)\cu_{,\eta}-1]\, ,\\
F_{\pm T_{\pm} \theta}&=&\Phi_{\pm,\theta}=\frac{\delta^{\pm1/2}}{\Delta_{\pm}}e^{2\cu}(1-\eta^2)\cu_{,\theta}\, .\n{2.9b}
\ea
 
\subsection{Singularities}

In this paper we mainly focus on study of the horizons ${\cal H}^{(\pm)}$, and the inner domain located between the horizons. Since one cannot trust the metric obtained by the analytical continuation of the exterior metric beyond the inner (Cauchy) horizon, it is reasonable to postpone study of the regions close to the spacetime singularity, until the classical and quantum (in)stability will be proved. For this reason we give only a couple of remarks about properties of the singularities in the analytic continuation of the charged distorted black hole solution.

The curvature and the electromagnetic field invariants diverge for $\Delta_{\pm}=0$, i.e. for
\ba\n{2.40}
\eta&=&-\frac{1+\delta_0e^{2\hu}}{1-\delta_0e^{2\hu}},
\ea
indicating the spacetime singularity. For RN black hole the singularity is located at $\eta=-1/p$, $p\in(0,1]$,  corresponding to $r=0$. Analyzing expression \eq{2.40} we see that for $\hu\leqslant 0$ the singularity is located in the region $\eta < -1$, whereas for $\hu > 0$ the space-time singularity is naked and located outside the outer horizon, $\eta > 1$. Thus, if the distortion field $\hu$ satisfies the strong energy conditions, i.e. $\hu\leqslant 0$, the spacetime outside the black hole outer horizon is regular, and the singularity is located behind the inner (Cauchy) horizon. 

\section{Duality relations between the inner and outer horizons}

In this Section we describe special symmetry relations between the inner and outer horizons. 
Consider a 2D subspace $T_{\pm}=const.$, $\phi=const.$ orthogonal to the corresponding Killing vectors. In the coordinates 
\be\n{psi}
\eta=\cos\psi\hhh \psi\in[0,\pi]
\ee
the subspace metric is
\be\n{sub}
d\Sigma^2_{\pm}=\Delta_{\pm}e^{-2\cu+2\hv}\left[-d\psi^2+d\theta^2\right].
\ee
Fig.~1 illustrates the Carter-Penrose diagram for these metrics. Lines $\psi\,\pm\,\theta=const.$ are null rays propagating from the outer to the inner horizon within the 2D subspace. One of such null rays is shown in the figure. It starts at point $A$ on the outer horizon ${\cal H}^{(+)}$, goes through the ``north pole" at $\theta=\pi$, and reaches point $B$ at the inner horizon ${\cal H}^{(-)}$.

\begin{figure}[htb]
\begin{center} 
\includegraphics[height=4.99cm,width=5cm]{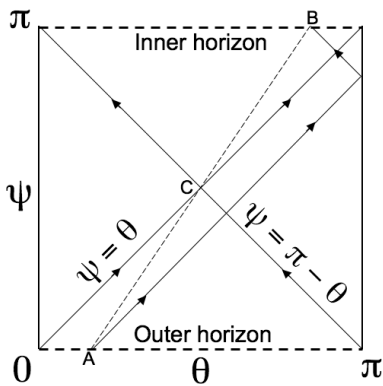} 
\caption{The Carter-Penrose diagram for $(\psi, \theta$) subspace of the charged distorted black hole interior. The arrows illustrate propagation of future directed null rays. Points $A$ and $B$ are symmetric with respect to the central point $C(\pi/2, \pi/2)$.} 
\end{center}\n{FF1}
\end{figure}

Consider a transformation $R_C$ representing the reflection of coordinates $(\psi,\theta)$ with respect to the `central point' $C$ in the interior region
\ba\n{2.10}
R_C: && (\psi, \theta)\to (\pi-\psi, \pi-\theta).
\ea
This transformation determines a map $R^*_C$ between functions  defined in the inner domain and on its boundaries
\be
f^*=R^*_C (f)\hh f^*(\psi, \theta)=f(\pi-\psi, \pi-\theta)\, .
\ee
Using the relations  \eq{2.6a}, \eq{PR}, \eq{v1}  and \eq{v2} we obtain
\ba
&&\cu^*(\psi,\theta)\equiv \cu(\pi-\psi,\pi-\theta)=\cu(\psi,\theta),\n{2.11a}\\
&&\hv^*_1(\psi,\theta)\equiv \hv_1(\pi-\psi,\pi-\theta)=-\hv_1(\psi,\theta)\, ,\n{2.11b}\\
&&\hv^*_2(\psi,\theta)\equiv \hv_2(\pi-\psi,\pi-\theta)=\hv_2(\psi,\theta)\, .\n{2.11c}
\ea

It is easy to see that the points $A$ and $B$ connected by a null ray (see Fig.~1) are related by the reflection $R_C$. Thus, the transformation $R^*_C$ determines a map between functions on the inner and outer horizons. Now we demonstrate that for \,$\cu$ and $\hv$ this is a symmetry transformation. In other words, the values of \,$\cu$ and $\hv$ on the inner horizon, $\psi=\pi$, are determined by their values on the outer horizon, $\psi=0$. 

Using \eq{2.11a}, \eq{2.6a}, \eq{PR} and the properties of the Legendre polynomials \eq{Lp}
we derive 
\ba\n{2.12a}
\cu(\pi,\pi-\theta)&=&\cu(0,\theta)=\sum_{n\geq 0}c_{n}\cos^n \theta-u_0.
\ea
Expressions \eq{PR}, \eq{v1}, \eq{v2} and \eq{Lp} give
\ba\n{v1a}
&&\hv_1(0,\theta)=-(1-\cos\theta)\sum_{n\geqslant 1}c_n\sum_{l=0}^{n-1}\cos^l\theta\nonumber\\
&&-(1+\cos\theta)\sum_{n\geqslant 1}(-1)^nc_n\sum_{l=0}^{n-1}(-\cos\theta)^l=2\,\cu(0,\theta),\nonumber\\
&&\hv_2(0,\theta)=0.
\ea
Thus, using \eq{2.11b} and \eq{2.11c} we have  
\ba\n{2.12b}
\hv(\pi,\pi-\theta)&=&-\hv(0,\theta)=-2\,\cu(0,\theta).
\ea
The above expressions \eq{2.12a} and \eq{2.12b} allow one to establish special symmetry relations between the geometric properties of the inner and outer horizons. We call relations \eq{2.12a}, \eq{2.12b} the {\em duality relations}.

Let us denote
\be\n{2.13}
\uu{\pm}(\theta)=\sum_{n\geq 0}(\pm 1)^{n}c_{n}\cos^n \theta-u_0.
\ee
As we shall see below, this function defines boundary values of the distortion fields, and as a result, the metric on the black hole horizons. It is easy to check that
\ba\n{2.14}
\uu{\pm}(\theta)&=&\uu{\mp}(\pi-\theta)\hhh \uu{\pm}(0)=\uu{\pm}(\pi)=0.
\ea
Expression \eq{2.13} implies that the functions $\uu{+}(\theta)$ and $\uu{-}(\theta)$ transform into each other under reflection with respect to the point $\theta=\pi/2$. This transformation property is directly related to the properties of the distortion field $\cu$. Namely, using \eq{2.12a}, \eq{2.14}, and \eq{2.6a}, \eq{PR} we derive the following {\em boundary values} of $\cu$
\ba
\cu(0,\theta)&=&\uu{+}(\theta)\hhh
\cu(\pi,\theta)=\uu{-}(\theta),\n{2.18}\\
\cu(\psi,0)&=&\uu{+}(\psi)\hhh
\cu(\psi,\pi)=\uu{-}(\psi).\n{2.19}
\ea
Analogously, using \eq{2.12b}, \eq{2.18}, \eq{2.14} and \eq{lf} we derive the {\em boundary values} of $\hv$
\ba
\hv(0,\theta)&=&2\uu{+}(\theta)\hhh \hv(\pi,\theta)=-2\uu{-}(\theta),\n{2.20a}\\
\hv(\psi,0)&=&0\hhh \hv(\psi,\pi)=0.\n{2.20b} 
\ea
Thus, the distortion fields calculated on the inner horizon are expressed through those calculated on the outer horizon. This fact allows one to make important conclusions about the distortion of the Cauchy horizon.

The boundary values of the distortion fields $\cu$ and $\hv$ define symmetry properties of the metrics on the black hole horizon surfaces. The surface of the outer and the inner horizon is defined by $T_{\pm}=const.$ and $\eta=\eta_{\pm}=\pm 1$, respectively. The corresponding dimensionless metrics derived from metric \eq{dmet} by applying \eq{psi} and the boundary conditions \eq{2.18} and \eq{2.20a} are
\be\n{2.21}
d\sigma^2_{\pm}=e^{\pm2u_\pm}d\theta^2+e^{\mp2u_\pm}\sin^2\theta d\phi^2.
\ee
The dimensional metrics on the horizon surfaces are (see \eq{2.1})
\be\n{2.21d}
d\sigma^{(\pm)2}=\Omega_{\pm}^2d\sigma^2_{\pm}.
\ee
Here, and in what follows $u_\pm\equiv u_\pm(\theta)$. The metric $d\sigma^2_+$ coincides with the metric on the distorted Schwarzschild black hole horizon surface \cite{FS}. The dimensionless areas of the horizon surfaces are
\be\n{ar}
{\cal A}^{(+)}={\cal A}^{(-)}=4\pi.
\ee
The metrics $d\sigma^2_+$ and $d\sigma^2_-$ are related to each other by the transformation 
\be\n{2.23}
u_+ \longleftrightarrow -u_-,
\ee
which according to \eq{2.13} implies the following {\em duality relations} between the outer and the inner horizons
\be\n{2.24}
c_{2n} \longleftrightarrow -c_{2n}\hhh c_{2n+1} \longleftrightarrow c_{2n+1}.
\ee
Thus, the metrics $dS_{\pm}^2$ are identical for distortions which have only odd multipole moments. The derived duality relations imply in particular that the inner (Cauchy) horizon of a distorted charged black hole solution obtained by the Harrison-Ernst transformation is regular, if the outer horizon is regular. This conclusion and its generalization to the case of rotating and charged black holes was proven recently in \cite{Hen1}, \cite{Hen2}.

\section{Gaussian Curvature}

In this section we discuss geometry of the distorted horizon surfaces. Gaussian curvature is a natural measure of intrinsic curvature of a 2D surface. It is equal to 1/2 of its scalar curvature. Gaussian curvature of a horizon surface was studied by several authors (e.g. \cite{GC1}- \cite{GC4}).   
For the metric \eq{2.21} the Gaussian curvature is given by
\be\n{2.22}
K_{\pm}=e^{\mp2u_{\pm}}\left[1\pm u_{\pm,\theta\theta}\pm3\cot\theta u_{\pm,\theta}-2u_{\pm,\theta}^2\right].
\ee
The dimensional Gaussian curvatures associated with metrics \eq{2.21d} are
\be\n{Gd}
K^{(\pm)}=\Omega_{\pm}^{-2}K_{\pm}.
\ee 

We shall illustrate our analysis of the charged distorted black hole considering simple examples of the lowest order multipole distortions. Namely, we shall consider quadrupole and octupole distortions for which the corresponding functions $\uu{\pm}$ read
\ba\n{2.25}
\uu{\pm}&=&-c_2\sin^2\theta\hhh \uu{\pm}=\mp c_3\sin^2\theta\cos\theta.
\ea
Here, $c_2$ and $c_3$ are the quadrupole and the octupole moments, respectively.

\begin{figure}[htb]
\begin{center}
\HR
\ba
&\hspace{-0.17cm}\includegraphics[height=3.86cm,width=4cm]{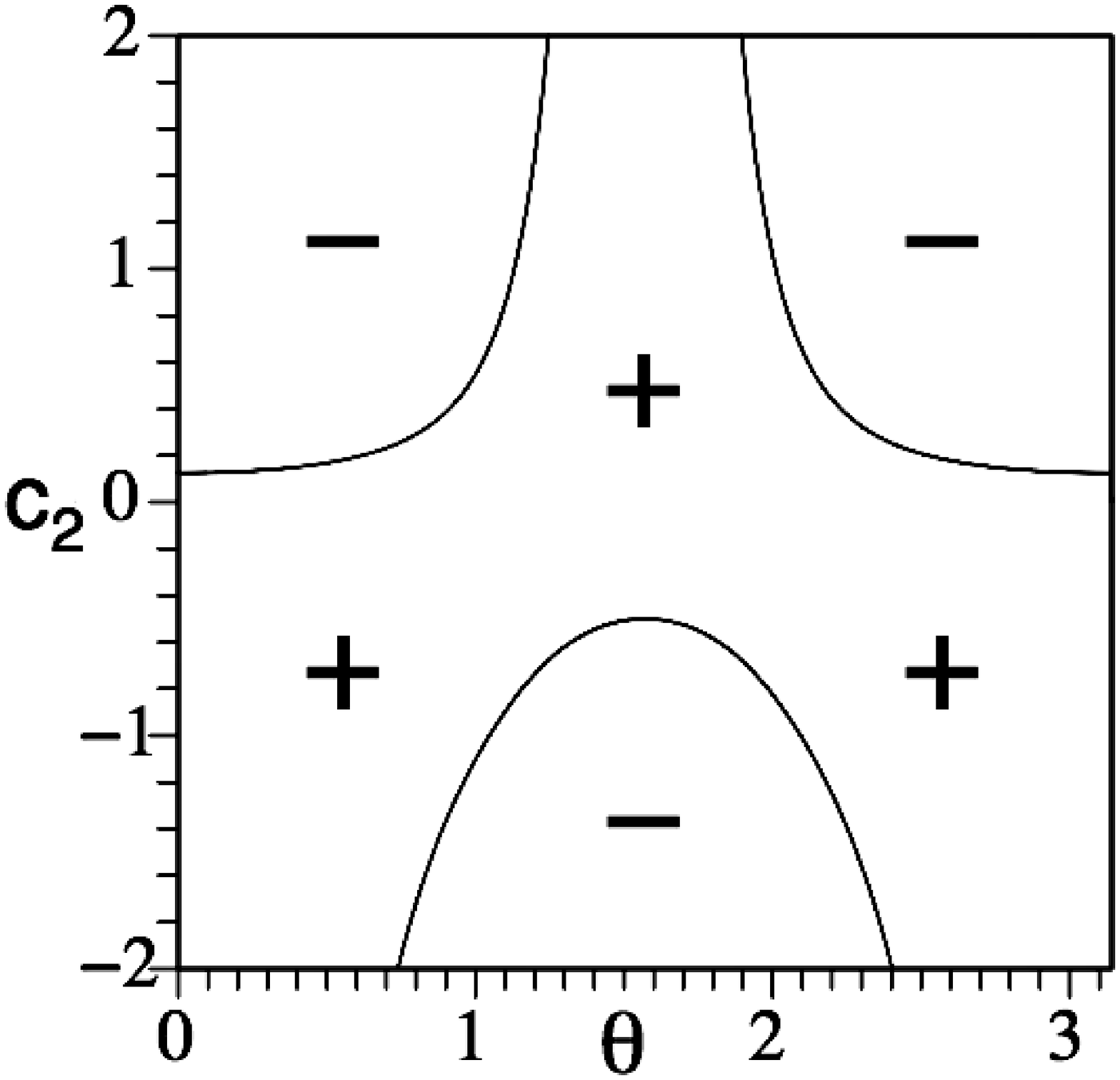}
&\hspace{0.37cm}\includegraphics[height=3.86cm,width=4cm]{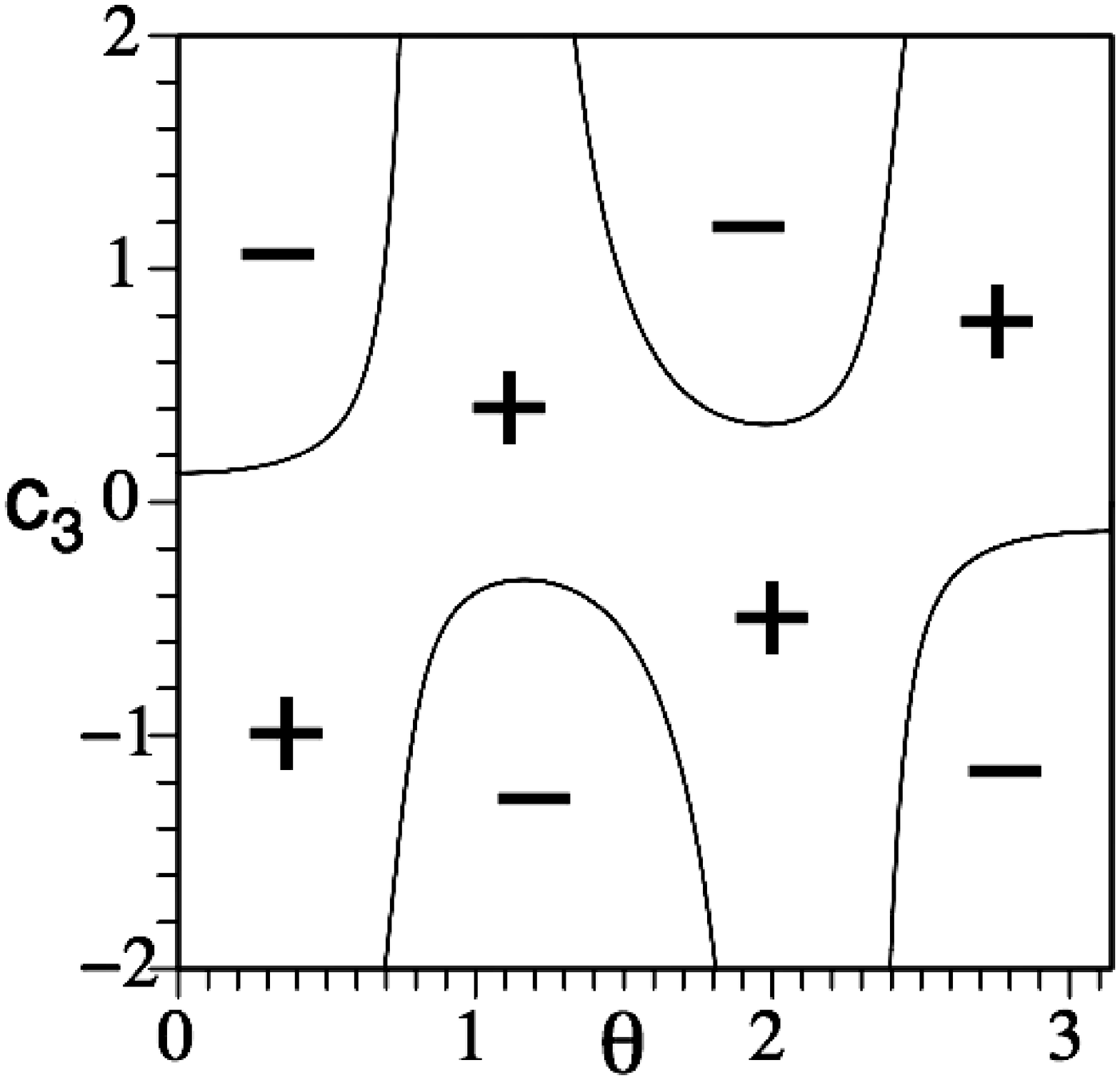}\nonumber\\
&\hspace{0.25cm}{\bf a} &\hspace{2.47cm}{\bf b}\nonumber
\ea
\vspace{-0.55cm}
\caption{Regions of positive and negative Gaussian curvature for the outer horizon surface. Plot ({\bf a}) illustrates the regions for different values of the quadrupole moment. Plot ({\bf b}) illustrates the regions for different values of the octupole moment. Curves separating these regions correspond to zero Gaussian curvature.}\label{f2}
\end{center} 
\end{figure}

Regions of positive and negative Gaussian curvature for different values of the quadrupole and octupole moments, for the outer horizon surface, are presented in Fig. 2. From the figure we see that for the quadrupole distortion regions of negative Gaussian curvature near the black hole poles ($\theta=0,\pi$) correspond to high positive values of $c_2$, and near its equator ($\theta=\pi/2$) to high negative values of $c_2$. Using \eq{2.22}, \eq{sumc} and the auxiliary expressions 
\ba\n{2.15}
\uu{\pm, \theta}(\theta)&=&-\sum_{n\geq 0}(\pm 1)^{n}c_{n}n\sin\theta\cos^{n-1}\theta,\\
\uu{\pm, \theta\theta}(\theta)&=&\sum_{n\geq 0}(\pm 1)^{n}c_{n}n\cos^{n-2}\theta[n\sin^2\theta-1],
\ea
we derive
\ba
K_{\pm}\vert_{\theta=0}&=&1\pm4\uuu{\pm}\hhh K_{\pm}\vert_{\theta=\pi}=1\pm4\uuu{\mp},\n{2.26}\\
K_{\pm}\vert_{\theta=\pi/2}&=&e^{\pm2(u_0-c_0)}(1\pm2c_2-2c_3^2).\n{2.27}
\ea
Here, 
\be\n{2.17}
\uuu{\pm}=-\sum_{n\geq 0}(\pm 1)^{n}c_{n}n.
\ee
Thus, the sign of the Gaussian curvature strictly depends on the distortion field. Using these expressions we derive that for the quadrupole distortion Gaussian curvature of the outer horizon surface is positive at the poles for $c_2<1/8$, and on the equator for $c_2>-1/2$. According to the duality relations \eq{2.24} regions of positive and negative Gaussian curvature of the inner horizon surface can be constructed by mirror reflection of Fig.~2 with respect to the line $c_2=0$.

\begin{figure}[htb]
\begin{center}
\HR
\ba
&\hspace{-0.23cm}\includegraphics[height=3.6cm,width=4cm]{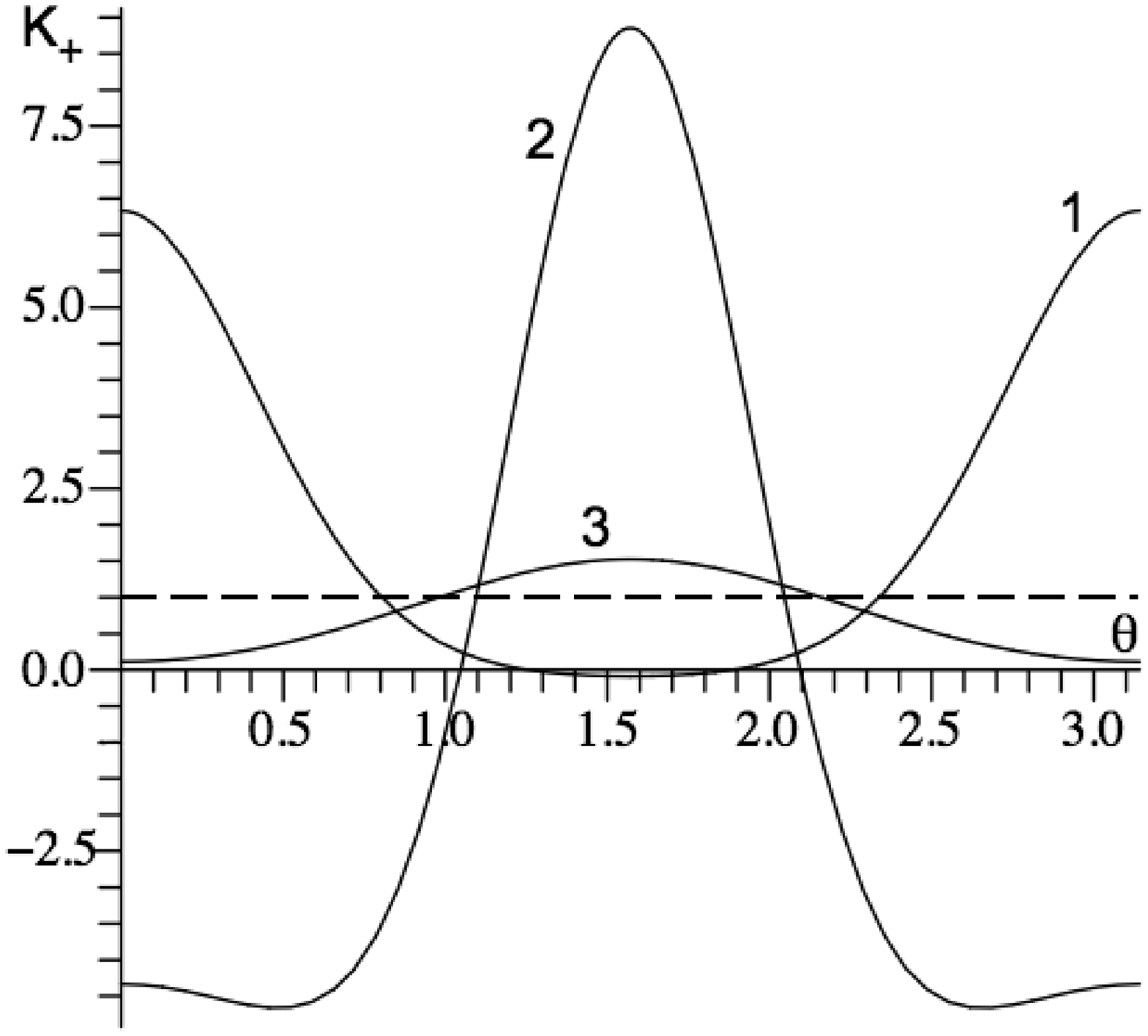}
&\hspace{0.27cm}\includegraphics[height=3.6cm,width=4cm]{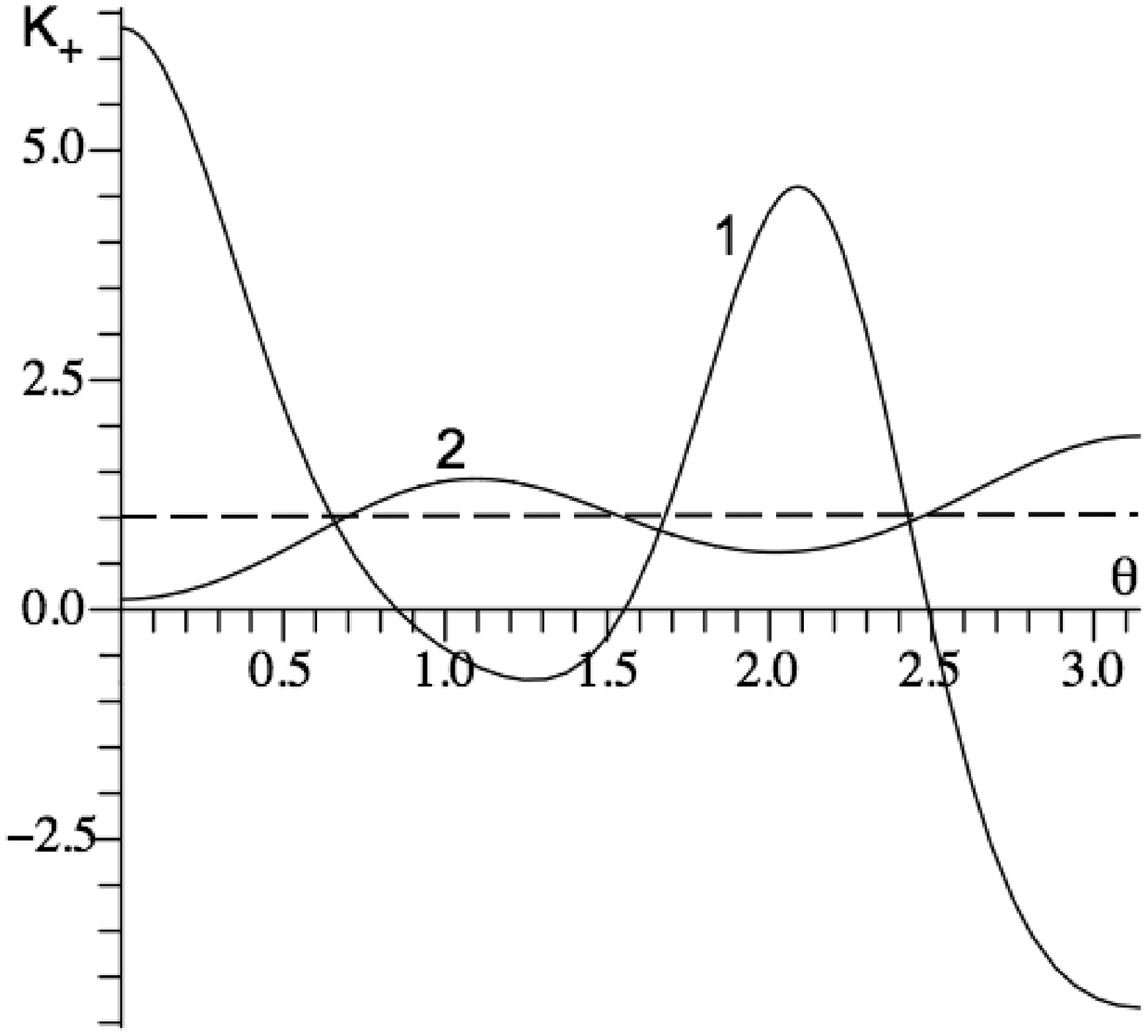}
\nonumber\\
&\hspace{0.1cm}{\bf a} &\hspace{2.32cm}{\bf b}\nonumber
\ea
\vspace{-0.55cm}
\caption{Dimensionless Gaussian curvature $K_+$ of the outer horizon surface. ({\bf a}) The quadrupole distortion: $c_{2}=-2/3$ (line 1), $c_{2}=2/3$ (line 2), and $c_{2}=1/9$ (line 3). ({\bf b}) The octupole distortion: $c_{3}=-2/3$ (line 1), and $c_{3}=1/9$ (line 2). Dashed horizontal lines of $K_+=1$ correspond to RN black hole.}\label{f3}
\end{center} 
\end{figure}

Fig. 2b illustrates that there is a symmetry between the regions of positive and negative Gaussian curvature and signs of the octupole moment. Namely, the transformation $c_3\to-c_3$, $\theta\to\pi/2-\theta$ leave the figure unchanged. Using \eq{2.26} we derive that for $c_3>1/8$ Gaussian curvature is negative on the ``north" pole and positive on the ``south" pole, whereas for $c_3<-1/8$ it is negative on the south pole and positive on the north. In addition, there are the regions of negative Gaussian curvature near the ``tropics" ($\pm23^{\circ}26'22''$ from the equator), i.e. near $\theta_{-}\approx1.165$ (corresponding to $\approx23^{\circ}16'39''$ from the equator) for $c_3<-0.333$, and $\theta_{+}\approx1.977$ (corresponding to $\approx-23^{\circ}16'39''$ from the equator) for $c_3>0.333$. According to the duality relations \eq{2.24} Gaussian curvature of the inner horizon surface is identical to that of the outer horizon surface. Dimensionless Gaussian curvature of the outer horizon surface for certain values of the quadrupole and octupole moments is plotted in Fig. 3.

As we shall see in Section VII, the curvature and the electromagnetic field invariants calculated on and at the vicinity of the black hole horizons are expressed in terms of the corresponding Gaussian curvatures and their derivatives. 

\section{Embedding}

To visualize the distorted horizon surfaces we present their isometric embedding into a flat 3D space. 
To construct the embedding we consider an axisymmetric 2D surface parametrized as follows
\ba\n{2.28}
\rho&=&\rho(\theta)\hhh z=z(\theta).
\ea
Let us embed this surface into a flat 3D space with the metric in cylindrical coordinates $(z,\rho,\phi)$:
\be\n{2.29}
dl^{2}=\epsilon dz^{2}+d\rho^2+\rho^2d\phi^2,
\ee
where for Euclidean space $\epsilon=1$, and for pseudo-Euclidean space $\epsilon=-1$ \cite{f3}.
The geometry induced on the surface is given by
\be\n{2.30}
dl^2=(\epsilon z_{,\theta}^2+\rho_{,\theta}^2)d\theta^2+\rho^2d\phi^2.
\ee
Matching metrics \eq{2.21} and \eq{2.30} we derive the following embedding map
\ba
\rho&=&e^{\mp\uu{\pm}}\sin\theta\hh
z=\int^{\pi/2}_{\theta} \mathcal{Z}\, d\theta\n{2.31a},\\
\mathcal{Z}^2&=&\epsilon e^{\pm2u_\pm}[1-e^{\mp 4u_\pm}(\cos\theta\mp u_{\pm,\theta}\sin\theta)^2]\n{2.31b}.
\ea
From \eq{2.31b} we see that if the expression in the square brackets is negative, an isometric embedding into 3D Euclidean space is not possible, and we should take $\epsilon=-1$.

\begin{figure}[htb]
\begin{center}
\HR
\ba
&\hspace{-0.03cm}\vspace{5cm}\includegraphics[height=4.06cm,width=4cm]{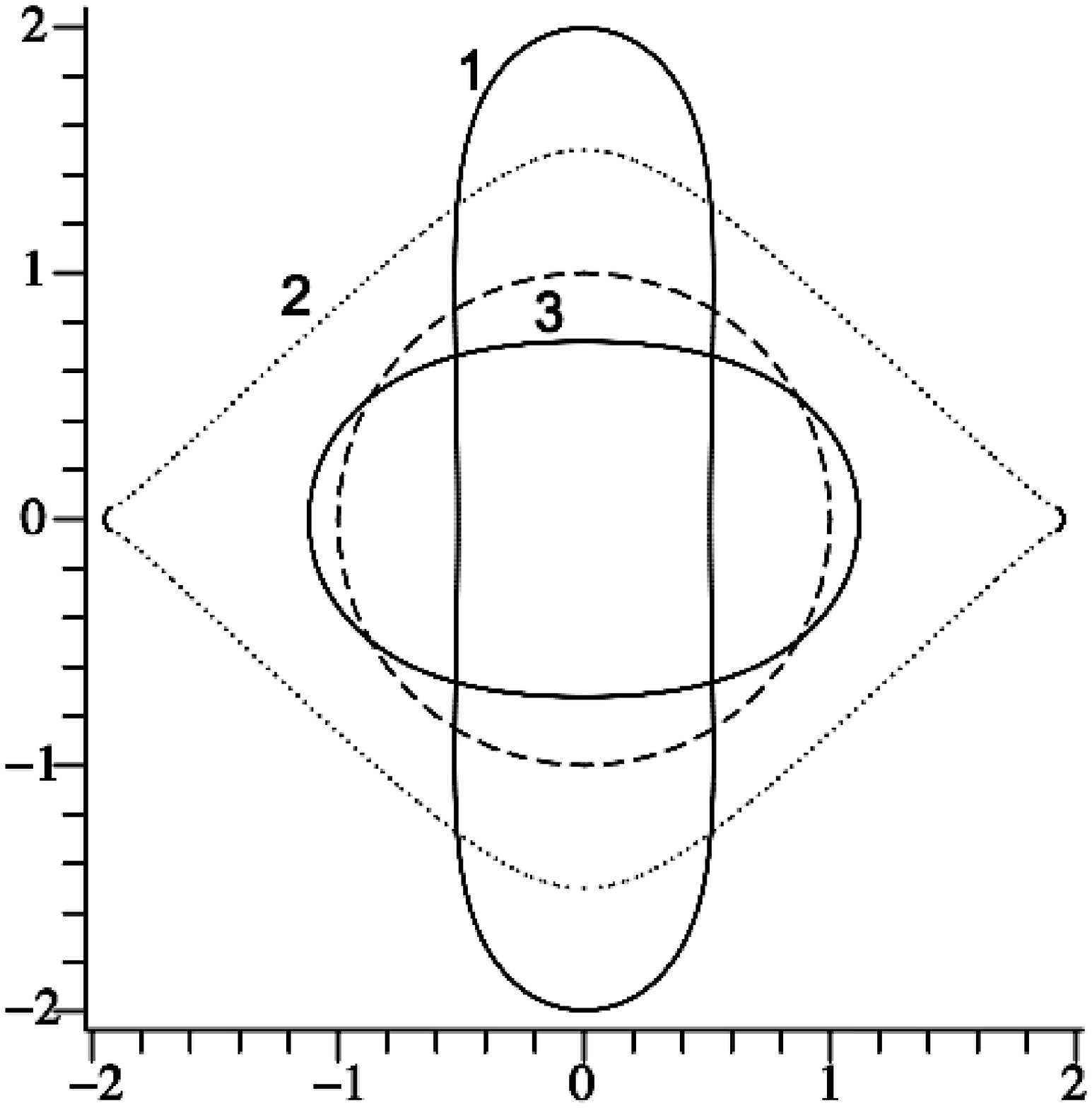}
&\hspace{0.21cm}\includegraphics[height=4.15cm,width=4cm]{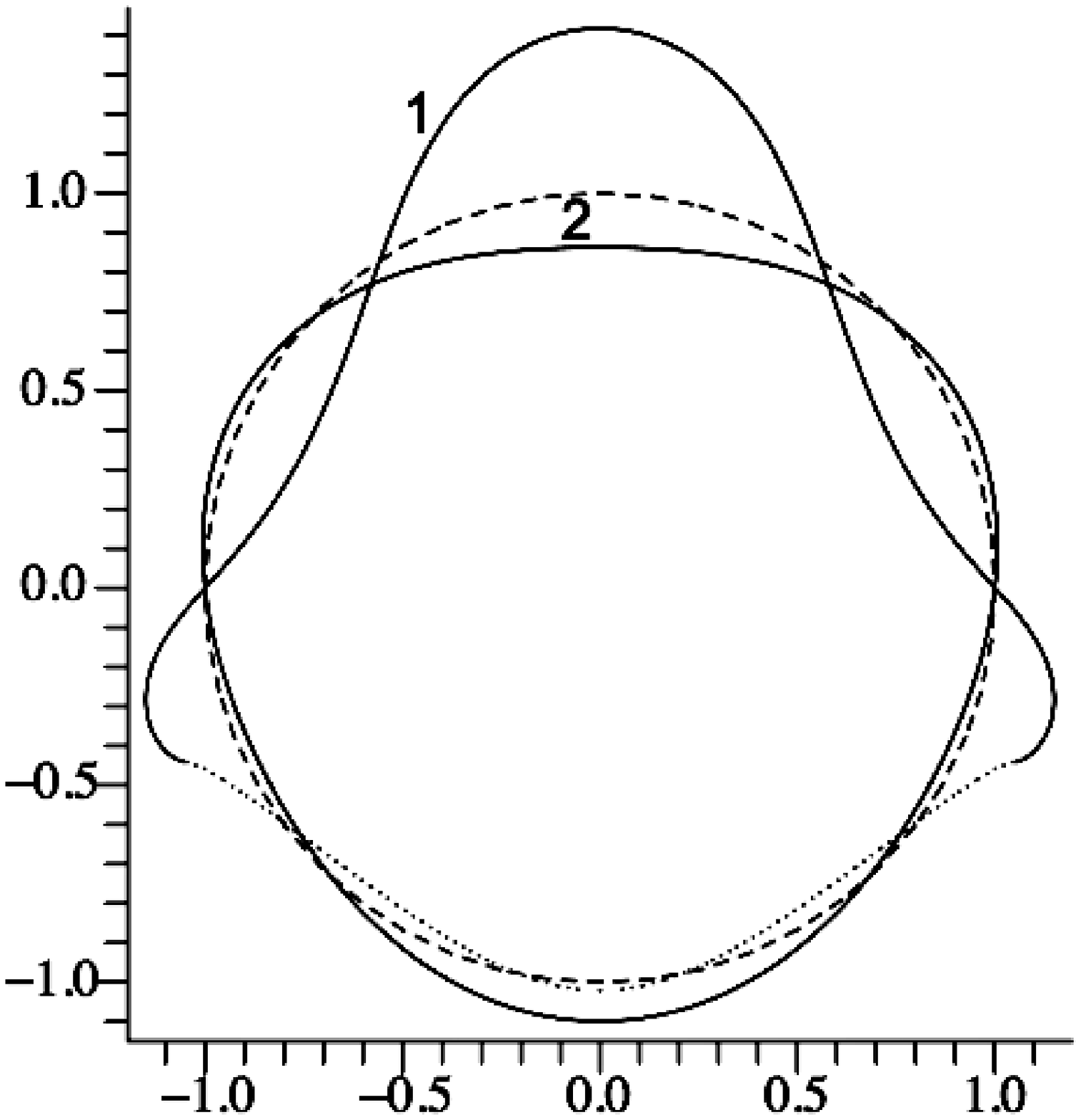}
\nonumber\\
&\hspace{0.1cm}{\bf a} &\hspace{2.32cm}{\bf b}\nonumber
\ea
\vspace{-0.55cm}
\caption{The shape of the outer horizon surface. The shape curves are shown in $(\rho,z)$ plane. ({\bf a}) The quadrupole distortion: $c_{2}=-2/3$ (line 1), $c_{2}=2/3$ (line 2), and $c_{2}=1/9$ (line 3). ({\bf b}) The octupole distortion: $c_{3}=-2/3$ (line 1), and $c_{3}=1/9$ (line 2). Regions embedded into pseudo-Euclidian space are illustrated by doted lines. Dashed circles of radius 1 correspond to RN black hole.}\label{f3a}
\end{center} 
\end{figure}

According to the duality relations \eq{2.24} it is enough to consider embedding of the outer horizon surface only. The shape curves of the outer horizon surface are presented in Fig. 4. The embedding diagrams for the outer horizon surface can be obtained by rotation of the curves around the vertical axis of symmetry lying in the plane of the figure,  parallel to $z$ axis. Note, that the change in sign from $`+`$ to $`-`$ of the quadrupole moment corresponds to deformation of the rotational curve from oblate to prolate and vice versa. This transformation corresponds to the duality relations \eq{2.24} between the outer and inner horizon surfaces. The change in sign of the octupole moment corresponds to overturn of the rotational curve preserving its shape.     

\section{Free fall from the outer to the inner horizon}

It is interesting to check how the distortion changes the maximal proper time of a free fall of a test particle from the outer to the inner horizon. Let us consider motion of a test particle of zero angular momentum which moves from the outer to the inner horizon along the axis of symmetry. Free fall from the north pole corresponds to $\theta=0$, and free fall from the south pole corresponds to $\theta=\pi$. We use metric \eq{2.1} with $dS_{+}^2$. Using \eq{lf} we derive the proper time of the free fall:
\be\n{time}
\tau(E)=\left. \Omega_+\int_{-1}^{+1}\frac{\Delta_+^{1/2}e^{-\cu}d\eta}{(\Omega_+^{-2}\Delta_+e^{-2\cu}E^2+1-\eta^2)^{1/2}}\right|_{\theta=0,\pi}\,,
\ee
where $E$ is the energy of the particle,
\be\n{E}
E=\Omega_+^2\frac{\eta^2-1}{\Delta_+}e^{2\cu}\frac{dT_+}{d\tau}.
\ee
The maximal proper time corresponds to $E=0$. Using the coordinate transformation \eq{psi} and applying \eq{2.19} we derive the maximal proper time for the free fall
\be
\tau_{max}=\tau(0)=\tau_+\Omega_+,
\ee
where the dimensionless time $\tau_+$ is
\be
\tau_+=\int^{\pi}_{0}\frac{d\psi}{2}\left[(\cos\psi+1)e^{-u(\psi)}-\delta e^{u(\psi)} (\cos\psi-1)\right].\\
\n{2.32} 
\ee
Here, $u(\psi)=u_+(\psi)$ for the fall from the north pole, and $u(\psi)=u_-(\psi)$ from the fall from the south pole. For RN black hole we have $\tau_+=\pi/(1+p)$, and $\tau_{max}=\pi M$, that is exactly the same as the maximal proper time for a free fall from event horizon to the singularity of Schwarzschild black hole of mass $M$ (\cite{MTW}, p. 836).

\begin{figure}[htb]
\begin{center}
\HR
\ba
&\hspace{-0.29cm}\includegraphics[height=4.05cm,width=4cm]{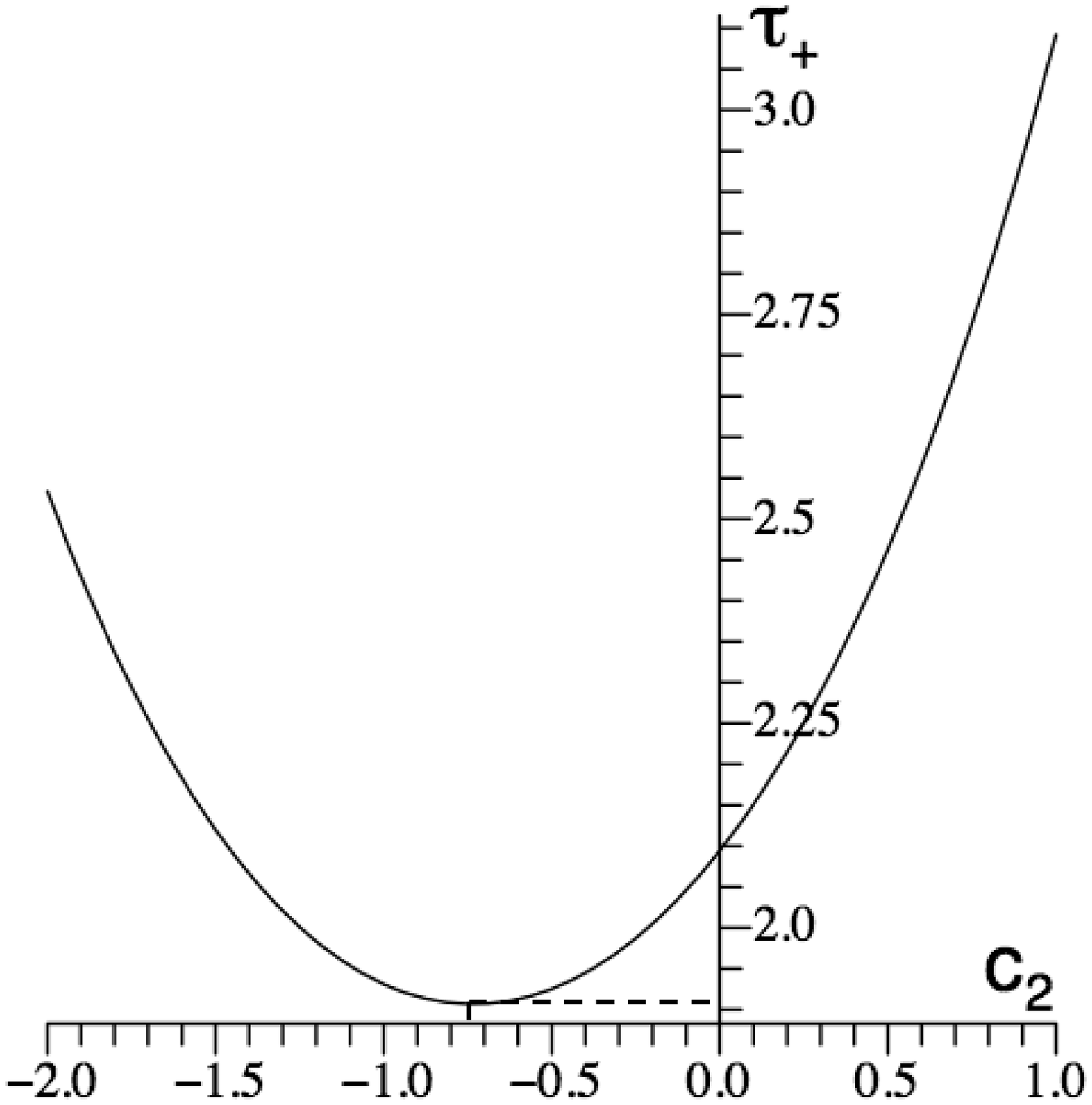}
&\hspace{0.24cm}\includegraphics[height=4.2cm,width=4cm]{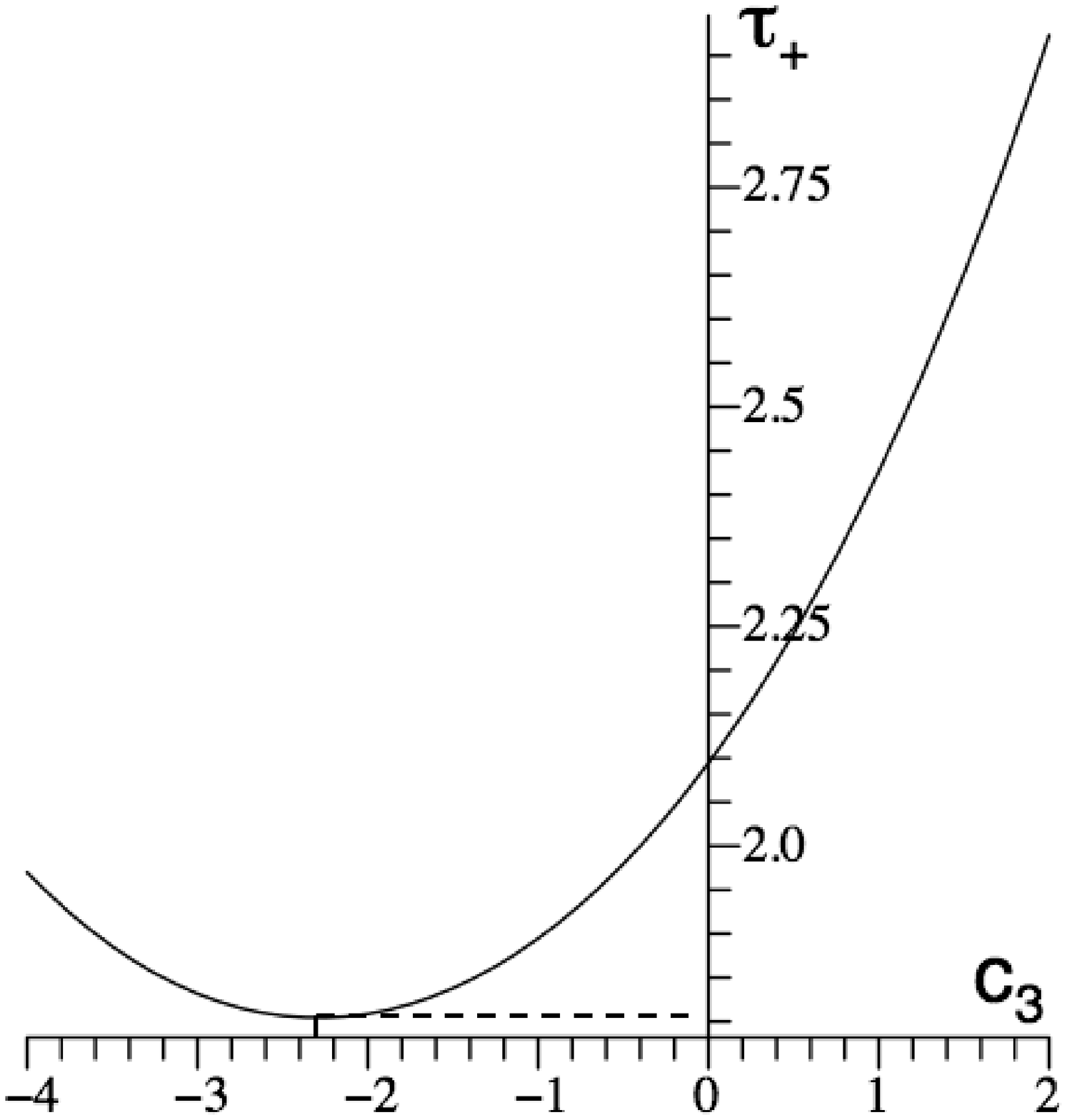}\nonumber\\
&\hspace{-0.3cm}{\bf a} &\hspace{2.15cm}{\bf b}\nonumber
\ea
\vspace{-0.55cm}
\caption{The free fall along the axis of symmetry from the outer to the inner horizon surface for $p'=1/2$. ({\bf a}) The dimensionless proper time $\tau_+$ for different values of the quadrupole moment $c_2$. Here, the minimal value of the dimensionless proper time $\tau_{+min}\approx 1.907$ corresponds to $c_{2min}\approx-0.734$. ({\bf b}) The dimensionless proper time $\tau_+$ for different values of the octupole moment $c_3$, for the fall from the north pole. Here, the minimal value of the dimensionless proper time $\tau_{+min}\approx 1.804$ corresponds to $c_{3min}\approx-2.292$, where $c_{3min}$ doesn't depend on the value of $p'$. For RN black hole $\tau_+=2\pi/3\approx2.094$.}\label{f4}
\end{center} 
\end{figure}

In the case of the quadrupole distortion \eq{2.25} the integral in \eq{2.32} can be calculated analytically:
\ba\n{2.33}
\tau_+&=&\frac{\pi}{2}I_0(c_2/2)[e^{c_2/2}+\delta e^{-c_2/2}],
\ea
where $I_0(x)$ is the modified Bessel function. Note, that because of the reflection symmetry of the horizon surfaces with respect to the plane $\theta=\pi/2$ the proper time is the same for the fall from the north and south poles. For the octupole distortion we evaluate the integral numerically. From expressions \eq{2.25} and \eq{2.32} we see that change in sign of the octupole moment corresponds to change of the poles as the starting points of the fall. The dimensionless proper time calculated for $p'=1/2$ is presented in Fig. 5  

\section{The Spacetime Invariants}

For distorted vacuum black holes there exists a remarkable relation between the Kretschmann scalar calculated on the surface of the event horizon ${\cal K}^{(+)}$ and the Gaussian curvature of the horizon $K^{(+)}$ calculated at the same point
\be\n{KKK}
{\cal K}^{(+)}=12{K^{(+)}}^2\, .
\ee
The proof of this relation can be found in \cite{FS}. This relation shows that the 4D curvature invariant of the spacetime calculated on the horizon is correlated with the shape of the horizon surface. In a region where the horizon is sharper the 4D curvature invariant is larger than in a region where the horizon is smoothed out. 
In order to prove the property \eq{KKK} one uses the fact that the horizon ${\cal H}^{(+)}$ surface is a {\em totally geodesic} surface. 
 
The general analysis by Boyer \cite{Boy}, and in particular his conclusion saying that a bifurcate Killing horizon contains a totally geodesic 2D surface, which is in fact independent of the field equations, can be applied to the case of the charged distorted black hole. For this reason one can expect the existence of a relation similar to \eq{KKK} and generalizing the latter. In this section we discuss this problem.

First of all, let us emphasize that in the presence of the electromagnetic field $F_{\alpha\beta}$ there exist an additional 4D invariant $F^2=F_{\alpha\beta}F^{\alpha\beta}$ characterizing the strength of the field. For the distorted black hole the calculations give the following value of this invariant on the outer horizon (see \eq{2.9}, \eq{2.9b}, \eq{dimf}, \eq{2.18}, and \eq{2.20a})
\be\n{f1h}
{F^{(+)}}^2=-{2\over M'^2}{(1-p')\over (1+p')^3}\, .
\ee
The minus sign on the right hand side reflects the fact that we are dealing with an electric (not magnetic) field. The Kretschmann scalar ${\cal K}$ and the Weyl invariant ${\cal C}^2$ are related as follows
\be\n{kr}
{\cal K}={\cal C}^2 +2(F^2)^2\, .
\ee
In the presence of matter, in order to characterize the `strength' of the gravitational field, it is more convenient to use the Weyl invariant. The calculations presented in appendix give for the Weyl invariant on the event horizon the following expression
\be\n{CCC}
{{\cal C}^2}^{(+)}=12 \left[{K^{(+)}}-{1\over 2}{F^{(+)}}^2\right]^2\, .
\ee
It is evident that in vacuum, when $F^2$ vanishes and the Kretschmann  invariant coincides with the Weyl invariant, this relation reduces to \eq{KKK}. The second term in the square brackets is constant on the horizon (see appendix and Eq. \eq{2.35} below). Hence, in the presence of the electrostatic field the Gaussian curvature of the horizon surface is, effectively, uniformly shifted by a positive value.

Similar relations are valid for the inner horizon
\ba\n{f2h}
{F^{(-)}}^2&=&-{2\over M'^2}{(1+p')\over (1-p')^3}\,,\\
{{\cal C}^2}^{(-)}&=&12 \left[K^{(-)}-{1\over 2}{F^{(-)}}^2\right]^2\, .
\ea
Using \eq{kr} we can calculate the ratio of the Kretschmann invariants on the black hole horizons:
\be\n{rat}
k=\frac{\mathcal{K}^{(+)}}{\mathcal{K}^{(-)}}=\delta^4\frac{3(K_++\delta)^2+2\delta^2}{3(K_-+\delta^{-1})^2+2\delta^{-2}}.
\ee 
This ratio calculated for $p'=1/2$ is presented on Fig. 6 below. The behavior of the curves is very similar to those for the Gaussian curvature illustrated on Fig. 3. 

\begin{figure}[htb]
\begin{center}
\HR
\ba
&\hspace{-0.29cm}\includegraphics[height=3.89cm,width=4cm]{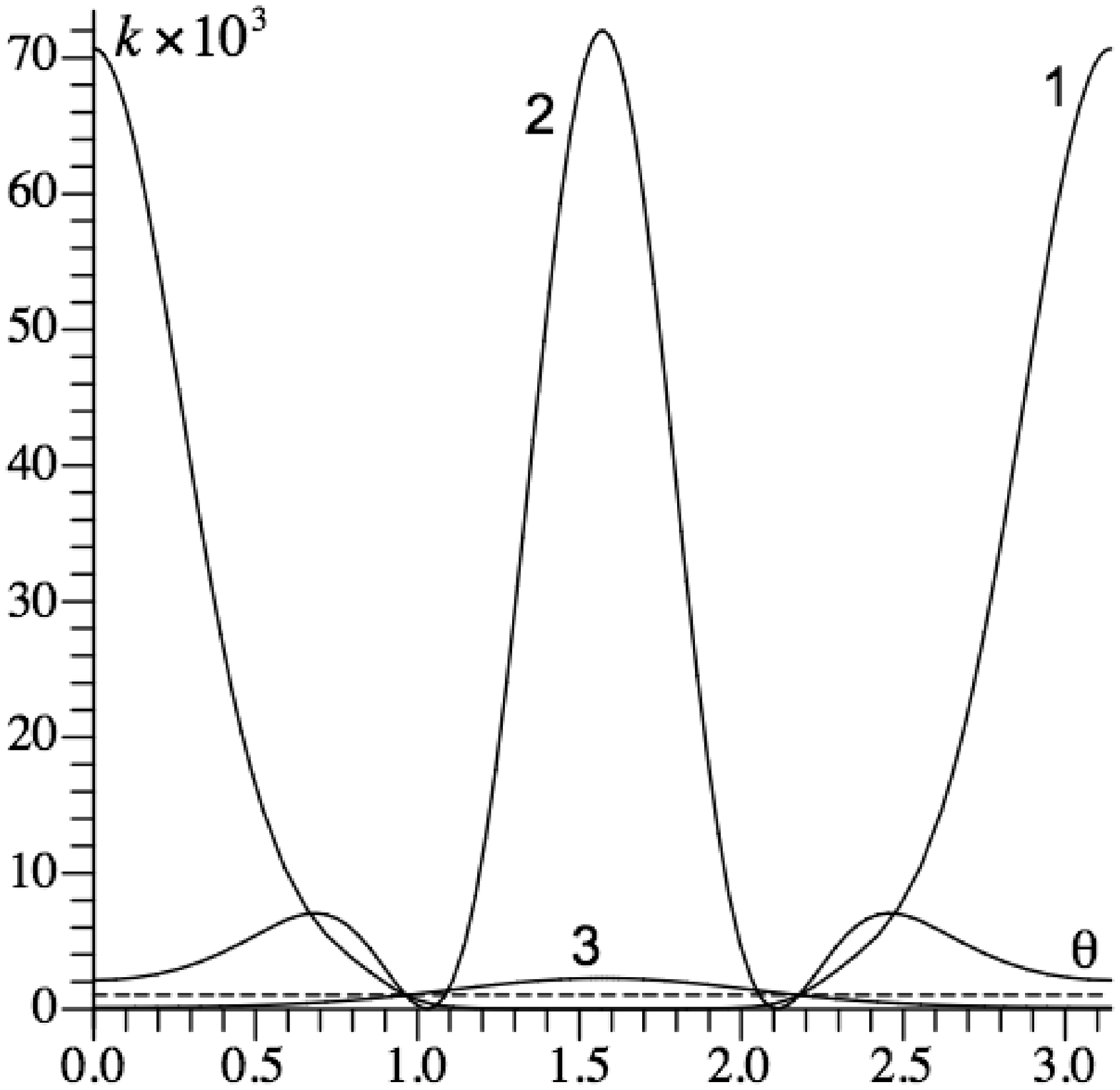}
&\hspace{0.24cm}\includegraphics[height=3.92cm,width=4cm]{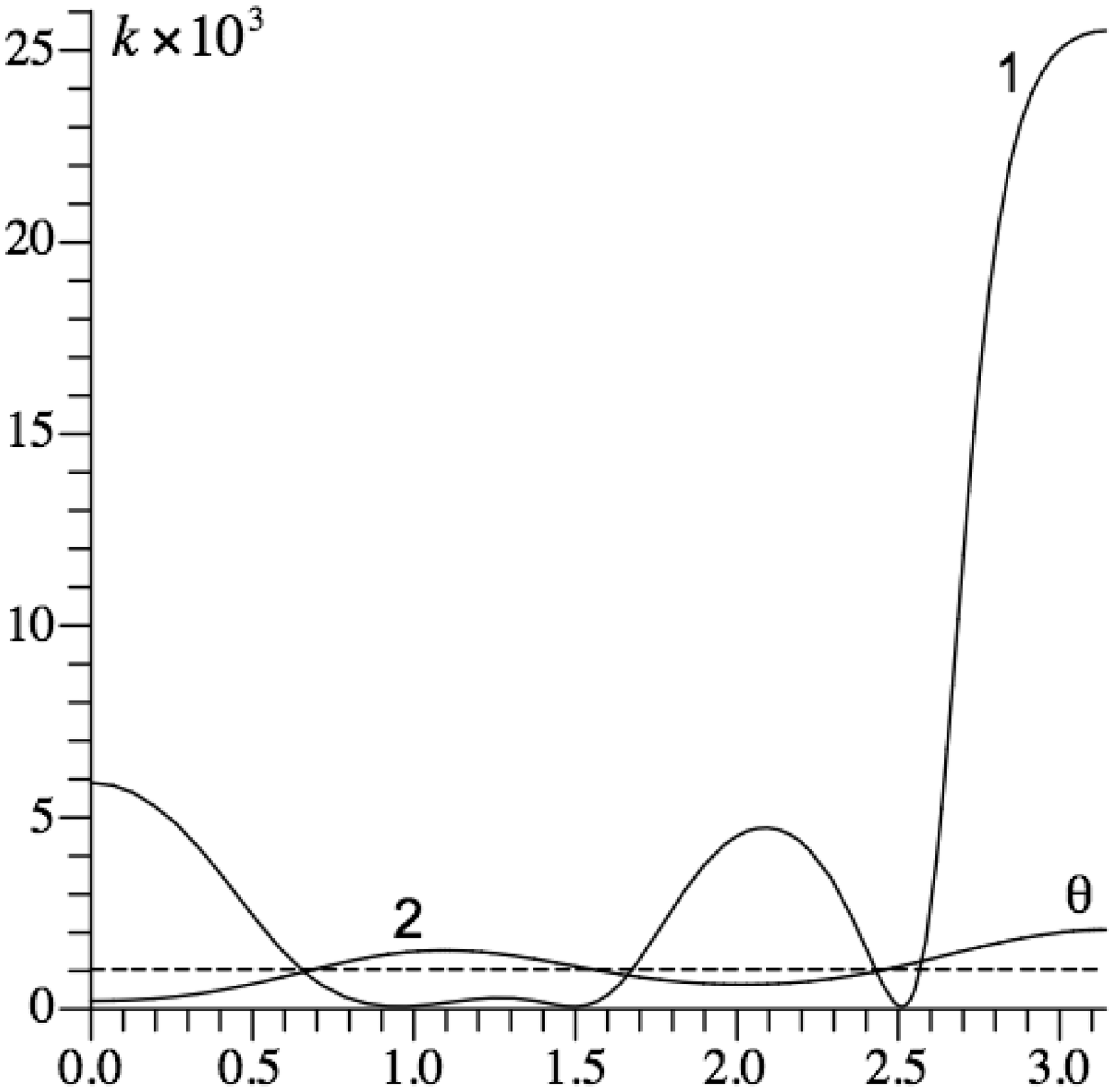}\nonumber\\
&\hspace{0.05cm}{\bf a} &\hspace{2.35cm}{\bf b}\nonumber
\ea
\caption{The ratio $k$ for $p'=1/2$. Plot ({\bf a}) illustrates the ratio for the quadrupole distortion of $c_2=-2/3$ (line 1), $c_2=2/3$ (line 2) and $c_2=1/9$ (line 3). Plot ({\bf b}) illustrates the ratio for the octupole distortion of $c_3=-2/3$ (line 1), and $c_3=1/9$ (line 2). The dashed horizontal line correspond to RN black hole.}
\label{f6} 
\end{center} 
\end{figure}

Finally, we present the expressions for the curvature and the electromagnetic field invariants at the vicinity of the black hole horizons. We use the results of the appendix \eq{1.17a}, \eq{1.14}, and \eq{1.17b}. The expansion of the electromagnetic field invariant near the black hole horizons reads
\be\n{2.35}
F_{\pm}^2=-2\delta^{\pm1}\pm4\delta^{\pm1}e^{\pm2u_{\pm}}(K_{\pm}-\delta^{\pm1})(\eta\mp1)+...\,.
\ee

The expansion of the Weyl invariant near the black hole horizons is
\ba\n{2.36}
{\cal C}^2_{\pm}&=&12K_{e\pm}^2\mp4\left(3K_{e\pm}^2[3K_{\pm}-2\delta^{\pm1}]e^{\pm2u_{\pm}}-2[K_{\pm,\theta}]^2\right.\nonumber\\
&+&\left.3K_{e\pm}[K_{\pm,\theta\theta}+\cot\theta K_{\pm,\theta}]\right)(\eta\mp1)+...\,,
\ea
where $K_{e\pm}=(K_{\pm}-\delta^{\pm1})$.

\section{Conclusion}

In this paper we studied interior of distorted, static, axisymmetric, electrically charged black hole. The corresponding metric was derived by the Harrison-Ernst transformation applied to the metric of distorted, static, axisymmetric vacuum black hole, whose interior was discussed in \cite{FS}. We established the special duality relations between the properties of the inner and outer horizons of the distorted charged black hole. These duality relations allow one to make a conclusion about the inner (Cauchy) horizon structure, which is based on the structure of the outer (event) horizon of the black hole. In particular, regions of positive and negative Gaussian curvature and its values on the outer horizon surface are correlated with those on the inner horizon surface. There is a correlation between the shapes of the horizon surfaces as well.

We derived expansion of the curvature and electromagnetic field invariants near the black hole horizons which is expressed in terms of the Gaussian curvature, electrostatic field and their derivatives calculated on the horizon surfaces. Thus, the established duality relations show that the spacetime geometry near the inner (Cauchy) horizon is correlated with the spacetime geometry near the outer (event) horizon. This implies that if the distortion leaves the outer horizon regular, the inner horizon remains regular as well.

The duality between the outer and inner horizons seems important. Apparently, according to the mass inflation phenomenon \cite{IP} such duality breaks in the case of dynamical perturbation of RN black hole. Namely, due to presence of the outgoing flux the inner apparent horizon and the Cauchy horizon become separated. The infinite grow of the mass parameter induced by the blue-shift of the ingoing flux on the Cauchy horizon is not canceled by the red-shift of the ingoing flux on the apparent horizon. As a result, the Cauchy horizon becomes singular. This doesn't happen in the case of static, axisymmetric distortion. One may think of the static distortion in the dynamical region between the black hole horizons as represented by standing waves. According to the duality relations between the horizons, initial and boundary values of the waves should be dual as well. 

Quite possibly, the axisymmetric, static distortion due to remote charged masses and fields can not affect much interior of the charged black hole. In such situation nothing enters, or leaves (through the Cauchy horizon into other ``universes") the black hole. Thus, the black hole inner horizon remains regular due to such type of distortion. Nevertheless, as our analysis shows, such ``serene" distortion can in fact deform interior of the black hole to create regions of high local curvature. Moreover, the distortion noticeably affects the maximal proper time of a free fall of a test particle moving along the axis of symmetry in the black hole interior. An important question if the Cauchy horizon of electrically charged black hole is regular for an arbitrary static, external distortion remains open.    

\begin{acknowledgments}

This research was supported  by the Natural Sciences and Engineering
Research Council of Canada and by the Killam Trust. We are grateful to A. J. S. Hamilton for his useful remarks. We also thank J. Hennig for bringing our attention to their papers \cite{Hen1}, \cite{Hen2}.

\end{acknowledgments}

\appendix

\section{\label{A} Calculation of the spacetime invariants near the black hole horizons}

In this appendix we obtain expressions for the curvature and electromagnetic field invariants near the black hole horizons. We start our construction in the regions where the Killing vector is timelike, namely outside of the horizons. Final expressions of the invariants will be valid in the region between the horizons as well.

The simplest curvature invariant is the Kretschmann scalar, which for Einstein-Maxwell 4D spacetime admits the following decomposition
\be\n{1.10}
\mathcal{K}=R_{\alpha\beta\gamma\delta}R^{\alpha\beta\gamma\delta}={\cal C}^2+2R_{\alpha \beta}R^{\alpha \beta},
\ee
where ${\cal C}^2=C_{\alpha \beta \gamma \delta}C^{\alpha \beta
\gamma \delta}$ is the Weyl scalar. The Weyl invariant characterizes
properties of a  `pure' gravitational field, while the square of the Ricci 
tensor $R_{\alpha \beta}R^{\alpha \beta}$ is
determined in our case by the electrostatic field. In this appendix we derive
an expansion of these invariants near the black hole horizons for an
arbitrary static, charged distorted black hole. In the main text of the
paper we shall use these results for a special case, when the static
spacetime is axisymmetric. A similar analysis for a vacuum distorted
black hole can be found in \cite{FrSa}. 

It is convenient to start with form of the metric proposed in
\cite{Israel}. Namely, we consider static spacetime and denote
timelike, hypersurface orthogonal Killing vector by $\BM{\xi}$. We assume that in the region
under consideration $\nabla_{\alpha}({\BM{\xi}}^2)$ does not vanish.
Following \cite{Israel} we write our metric, $g_{\alpha\beta}$
($\alpha,\beta, \ldots =0,\ldots ,3$) in this region in the form
\ba\n{1.2}
ds^2&=&-k^2dt^2+d\gamma^2\hhh d\gamma^2=\gamma_{AB}dy^A dy^B\\
&=&\kappa^{-2}(k,\theta^c)dk^2+h_{ab}(k,\theta^c)d\theta^ad\theta^b\,  .\nonumber
\ea
Here,  $k=(-\xi_{\alpha}\xi^{\alpha})^{1/2}$; $A, B, \ldots =1,2,3$; $a,b,c,\ldots=2,3\,$,
\be
\kappa^2=-\frac{1}{2}(\nabla^{\beta}\xi^{\alpha})(\nabla_{\beta}\xi_{\alpha})\, ,
\ee
and $h_{ab}$ is the metric on `equipotential' 2D surfaces
$k=const.$ spanned by $\theta^a$ coordinates.  At the horizon of a
static black hole, that is for $k=0$, $\kappa$ coincides with the
surface gravity. In a static spacetime the Weyl invariant can be
written as follows \cite{FZ} 
\ba\n{C2}
{\cal C}^2&\equiv&
C_{\alpha\beta\gamma\delta}C^{\alpha\beta\gamma\delta}=
8\Pi_{\alpha\beta}\Pi^{\alpha\beta}+8\Pi_{\alpha\beta}\Lambda^{\alpha\beta}\nonumber\\
&+&4\Lambda_{\alpha\beta}\Lambda^{\alpha\beta}-(\Pi+\Lambda)^2-2R_{\alpha\beta}R^{\alpha\beta}\, ,
\ea
where
\ba
\Pi_{\alpha\beta}&=&R_{\alpha\gamma\delta\beta}\zeta^{\gamma\delta}\hhh
\Pi\equiv\Pi_{\alpha}^{\,\,\,\alpha}=-\zeta_{\alpha\beta} R^{\alpha\beta}\, ,\\
\Lambda_{\alpha\beta}&=&R_{\alpha\beta}+\zeta_{\alpha\beta}\Pi\hhh
\Lambda\equiv\Lambda_{\alpha}^{\,\,\,\alpha}=R+\Pi\, .
\ea 
Here $\zeta_{\gamma\delta}=-\xi_{\gamma}\xi_{\delta}/\BM{\xi}^2$. For a
static spacetime $\Pi_{00}=\Pi_{0A}=0$. To calculate ${\cal C}^2$ it is
convenient to use the Gauss-Codazzi equations  
\ba\n{1.3g}
&&\hspace{-1cm}R_{ABCD}={\cal R}_{ABCD}+\varepsilon[{\cal S}_{AD}{\cal S}_{BC}-{\cal S}_{AC}{\cal S}_{BD}]\,,\\
&&\hspace{-1cm}n^{\alpha}R_{\alpha B C D}={\cal S}_{BC|D}-{\cal S}_{BD|C}\, ,\\
&&\hspace{-1cm}k R_{A\gamma\delta B}n^{\gamma}n^{\delta}=
-k \Pi_{AB}\nonumber \\
&&\hspace{-1cm}=\gamma_{AC}{\cal S}_{B\,\,\,,t}^{\,\,\,C}
+\varepsilon k_{|AB}+k {\cal S}_{AC}{\cal S}_{B}^{\,\,\,C}\,
.\n{1.3g_a}
\ea  
Here $n^{\alpha}=\xi^{\alpha}/k$ is the unit normal to hypersurface
$t=const.$, $\varepsilon=\BM{n}^2=-1$, ${\cal S}_{AB}$ is the
extrinsic 3D curvature of a hypersurface $t=const.$, ${\cal R}_{ABCD}$ is its
3D intrinsic curvature defined with respect to the metric
$d\gamma^2$, while ${\cal R}$ is the 3D scalar curvature. The stroke stands for a covariant
derivative with respect to this metric.

Relations \eq{1.3g}-\eq{1.3g_a} imply
\ba\n{1.3j}
&&\hspace{-1cm}2G_{\alpha\beta}n^{\alpha}n^{\beta}=-\varepsilon
{\cal R}-{\cal S}_{AB}{\cal S}^{AB}+{\cal S}^2\, ,\\
&&\hspace{-1cm}R_{\alpha\beta}n^{\alpha}n^{\beta}=-{\cal S}_{AB}{\cal S}^{AB}-\varepsilon k^{-1}k_{|A}^{\,\,\,\,\,|A}-k^{-1}{\cal S}_{,t}\, ,\\
&&\hspace{-1cm}G_{\alpha B}n^{\alpha}=R_{\alpha B}n^{\alpha}=
-{\cal S}_{,B}+{\cal S}_{B\,\,\,|C}^{\,\,\,C}\, ,\\
&&\hspace{-1cm}R_{AB}={\cal R}_{AB}-\varepsilon {\cal S}{\cal S}_{AB}
-k^{-1}k_{|AB}-\varepsilon k^{-1}\gamma_{AC}{\cal S}_{B\,\,\,,t}^{\,\,\,C}\,.\nonumber\\
\n{b14}
\ea 
Here ${\cal S}=\gamma^{AB}{\cal S}_{AB}$ is twice the mean curvature.
Since metric \eq{1.2} is static, the extrinsic curvature defined as
\be
{\cal S}_{AB}=\frac{1}{2}k^{-1}\gamma_{AB,t}
\ee
vanishes. Thus, \eq{1.3g}-\eq{b14} imply
\ba
&&\hspace{-1cm}\Pi_{AB}=k^{-1}k_{|AB}\hhh\Pi=k^{-1}k_{|A}^{\,\,\,\,\,|A}\, ,\\
&&\hspace{-1cm}\Lambda_{AB}={\cal R}_{AB}-k^{-1}k_{|AB}\hhh \Lambda = {\cal R}-k^{-1}k_{|A}^{\,\,\,\,\,|A}\, ,\\
&&\hspace{-1cm}\Lambda_{00}=0\hhh \Lambda_{0A} = 0\, .
\ea
The Einstein equations $G_{\alpha\beta}=8\pi T_{\alpha\beta}$ give
\ba\n{ein}
{\cal R}&=&16k^{-2}\pi T_{00}\hhh T_{0A}=0,\nonumber\\
G_{AB}&=&8\pi T_{AB}+k^{-1}k_{|AB}-k^{-1}\gamma_{AB}k_{|A}^{\,\,\,\,\,|A}.
\ea 
Thus, the Weyl invariant \eq{C2} written in terms of 3D objects related to
hypersurface $t=const.$ is 
\ba\n{Kr}
{\cal C}^2&=&2k^{-2}\left(k_{|AB}k^{|AB}-3k_{|A}^{\,\,\,\,\,|A}k_{|B}^{\,\,\,\,\,|B}\right)\nonumber\\
&+&2\left({\cal R}_{AB}+2k^{-1}k_{|AB}\right){\cal R}^{AB} \, .
\ea

The next step is a $(2+1)$-decomposition. We use the following expression for the 3D metric
\be\n{split}
d\gamma^2=\kappa^{-2}(k,\theta^c)dk^2+h_{ab}(k,\theta^c)d\theta^ad\theta^b\,.
\ee
We denote a covariant derivative with respect to the 2D-metric $h_{ab}$ as $(\ldots)_{:a}$. A unit vector orthogonal to equipotential 2D surface $k=const.$ is $n^{A}=\kappa\delta^{A}_{\,\,\,k}$, $\varepsilon=\BM{n}^2=1$. The extrinsic curvature of the surface is
\be
S_{ab}=\frac{\kappa}{2}h_{ab,k}.
\ee   
Using \eq{split} we derive
\ba\n{der}
k_{|kk}&=&\kappa^{-1}\kappa_{,k}\hhh k_{|ka}=\kappa^{-1}\kappa_{:a}\hhh k_{|ab}=\kappa S_{ab},\nonumber\\
k_{|A}^{\,\,\,\,\,|A}&=&\kappa S+\kappa\kappa_{,k}\hhh S=h^{ab}S_{ab}.
\ea
To project the Einstein equations on the $2D$ surface we have to define the stress-energy tensor of the electrostatic field. The electrostatic potential is given by $\Phi=\Phi(k,\theta^a)$. The corresponding electric field vector defined with respect to Schwarzschild time $t$ on hypersurface $t=const.$ reads 
\be\n{1.4}
E_{A}=-k^{-1}F_{0A}=-k^{-1}\Phi_{,A}. 
\ee 
We are interested in deformation of equipotential 2D surfaces. Thus, it is convenient to define orthogonal to the surfaces component of the electric field vector separately. The electric field vector components in an orthonormal frame are 
\be\n{1.5}
E_{\hat{k}}=-\kappa\,k^{-1}\Phi_{,k}\hhh E_{a}=k^{-1}\Phi_{:a}\, . 
\ee   
Thus, in our notations
\be\n{1.6} 
\Vec{E}^2=E_{\hat{k}}^2+k^{-2}\Phi_{:a}\Phi^{:a}. 
\ee 
The energy momentum tensor of the field is 
\be\n{1.7} 
8\pi T_{\alpha\beta}=2\xi_{\alpha}\xi_{\beta}k^{-2}\Vec{E}^2-2E_{\alpha}E_{\beta}+g_{\alpha\beta}\Vec{E}^2.
\ee
Using relations \eq{1.3j}-\eq{b14} for metric \eq{split} together with \eq{ein} we derive the Einstein equations projected onto 2D equipotential surfaces:  
\ba
\kappa^3\,S_{a\,\,\,,k}^{\,\,b}&=&\kappa^2[K-E_{\hat{k}}^2-k^{-2}\Phi_{:c}\Phi^{:c}]\delta_a^{\,\,b}-\kappa^3k^{-1}S_a^{\,\,b}\nonumber\\
&+&\kappa\kappa_{:a}^{\,\,:b}-2\kappa_{:a}\kappa^{:b}-\kappa^2
SS_a^{\,\,b}+2\kappa^2k^{-2}\Phi_{:a}\Phi^{:b},\nonumber\\
\n{1.8a}\\
\kappa^3\,S_{,k}&=&\kappa^2\left[\kappa
k^{-1}S-S_a^{\,\,b}S_b^{\,\,a}-2k^{-2}\Phi_{:a}\Phi^{:a}\right]\nonumber\\
&+&\kappa\kappa_{:a}^{\,\,:a}-2\kappa_{:a}\kappa^{:a},\n{1.8b}\\
k^{-1}\kappa\,\kappa_{,k}&=&-\kappa
k^{-1}S+E_{\hat{k}}^2+k^{-2}\Phi_{:a}\Phi^{:a},\n{1.8c}\\
\kappa^2\,E_{\hat{k}\,,k}&=&-\kappa
SE_{\hat{k}}-\kappa_{:a}k^{-1}\Phi^{:a}+\kappa\,k^{-1}\Phi_{:a}^{\,\,:a}.\n{1.8d}
\ea 
The corresponding constraints are 
\ba
0&=&S^2-S_a^{\,\,b}S_b^{\,\,a}-2K+2[\kappa
k^{-1}S+E_{\hat{k}}^2-k^{-2}\Phi_{:a}\Phi^{:a}],\nonumber\\
\n{1.9a}\\
0&=&[S_{:a}-S_{a\,\,\,:b}^{\,\,b}]k+2E_{\hat{k}}\Phi_{:a}+\kappa_{:a}.\,\,\,\,\,\,\n{1.9b}
\ea
Here, $K$ is the Gaussian curvature of a 2D equipotential surface $k=const.$

The square of the Ricci tensor $R_{\alpha \beta}R^{\alpha \beta}$ is equal to the squared electromagnetic field invariant
\be\n{1.11a}
R_{\alpha \beta}R^{\alpha \beta}=(F^2)^2=(F_{\alpha\beta}F^{\alpha\beta})^2.
\ee 
According to \eq{1.4} and \eq{1.6} $F^2$ has the following form
\be\n{1.11b}
F^2=-2\Vec{E}^2=-2[E_{\hat{k}}^2+k^{-2}\Phi_{:a}\Phi^{:a}].
\ee
Using expressions \eq{ein}, \eq{1.5}-\eq{1.8d} we derive
\ba\n{A1}
{\cal R}_{kk}&=&k^{-1}\kappa^{-1}\kappa_{:k}+\kappa^{-2}[\Vec{E}^2-2E^2_{\hat{k}}],\nonumber\\
{\cal R}_{ak}&=&k^{-1}\kappa^{-1}[\kappa_{:a}+2E_{\hat{k}}\Phi_{:a}],\nonumber\\
{\cal R}_{ab}&=&k^{-1}\kappa S_{ab}+h_{ab}\Vec{E}^2-2k^{-2}\Phi_{:a}\Phi_{:b},\nonumber\\
{\cal R}&=&2\Vec{E}^2=2k^{-1}\kappa[S+\kappa_{:k}].
\ea

Using \eq{Kr}, \eq{1.10}, \eq{1.11b} and expressions \eq{der} and \eq{A1} we have
\ba\n{1.12}
\frac{1}{8}{\cal C}^2&=&[\kappa^2S_a^{\,\,b}S_b^{\,\,a}+2\kappa_{:a}\kappa^{:a}+\kappa^2S^2+2\Vec{E}^2\Phi_{:a}\Phi^{:a}]k^{-2}\nonumber\\
&+&4E_{\hat{k}}\kappa_{:a}\Phi^{:a}k^{-2}-2\kappa\,[S_a^{\,\,b}\Phi_{:b}+S\Phi_{:a}]\Phi^{:a}k^{-3}.\nonumber\\
\ea

The hypersurface orthogonal Killing vector field $\xi^{\alpha}$ by definition is null on Killing horizon which is bifurcate ($\kappa\ne0)$. A bifurcate Killing horizon contains 2D spacelike, totally geodesic surface \cite{Boy}. In our coordinates this equipotential surface is defined by $t=const.$ and $k=0$. On the other side, a necessary and sufficient condition that a hypersurface is totally geodesic is its vanishing extrinsic curvature defined in the corresponding enveloping space \cite{Eis}. Thus, for the equipotential surfaces $t=const.$, $k=0$ we have $S_{ab}=0$. For a regular horizon its 2D surface has everywhere finite Gaussian curvature, and the electrostatic field on the surface is finite as well. Thus, we can deduce from the constraints \eq{1.9a}, \eq{1.9b} that on the horizon $\Phi_{:a}=\kappa_{:a}=0$. Hence, the electrostatic field potential $\Phi$ and the surface gravity $\kappa$ are constant on the horizon, as it has to be for a static black hole. This is nothing but the zeroth law of black hole thermodynamics \cite{4LBHM}. 

Projecting the first \eq{1.8a}, and the second \eq{1.8b} of the Einstein equations on the horizon, and using the first constraint \eq{1.9a} we derive
\be\n{1.13}
\left.2\kappa S_a^{\,\,b}k^{-1}\right\vert_{\hnH}=\left.\delta_a^{\,\,b}[K-E_{\hat{k}}^2]\right\vert_{\hnH}.
\ee
Here and below $(...)\vert_{H}$ means calculated on the horizon. Thus, from \eq{1.11a}, \eq{1.11b} and \eq{1.12} we derive the following expressions for the spacetime invariants calculated on the horizon
\ba\n{1.14}
\left.F^4\right\vert_{\hnH}&=&\left.R_{\alpha \beta}R^{\alpha \beta}\right\vert_{\hnH}=\left.4E_{\hat{k}}^4\right\vert_{\hnH},
\ea
and
\ba\n{1.15}
\left.{\cal C}^2\right\vert_{\hnH}&=&\left.12[K-E_{\hat{k}}^2]^2\right\vert_{\hnH}.
\ea  
This expression generalizes the relation between Gaussian curvature and the Kretschmann scalar calculated on the event horizon surface of an arbitrary distorted Schwarzschild black hole \cite{FS}, \cite{FrSa}.
  
We can expand the metric and the electrostatic field in series near the horizon and substituting these expansions into \eq{1.11b} and \eq{1.12} derive expressions of the spacetime invariants near the horizon. There are two types of quantities, even and odd in $k$, which we denote by $A=\{\kappa,h_{ab},K,\Phi,E_{\hat{k}}, F^2, {\cal C}^2\}$ and $B=\{S_a^{\,\,b},S\}$, respectively. The series expansions of $A$ and $B$ read 
\ba\n{1.16}
A&=&\sum_{n\geqslant 0}A^{[2n]}k^{2n}\hhh B=\sum_{n\geqslant 0}B^{[2n+1]}k^{2n+1}. 
\ea
The first term in $A$ gives its value on the horizon.  We can express higher order coefficients in the expressions in terms of these on the horizon substituting \eq{1.16} into the Einstein equations \eq{1.8a}-\eq{1.9b}. The necessary coefficients to calculate the first order expansion of the spacetime invariants are the following 
  
\ba\n{B1}
\kappa^{[2]}&=&\frac{1}{2\kappa^{[0]}}[2E_{\hat{k}}^{[0]2}-K^{[0]}]\hhh 
\Phi^{[2]}=-\frac{E_{\hat{k}}^{[0]}}{2\kappa^{[0]}},\nonumber\\
\nonumber\\
S_a^{\,\,b[1]}&=&\frac{\delta_a^{\,\,b}}{2\kappa^{[0]}}[K^{[0]}-E_{\hat{k}}^{[0]2}]\hhh 
S^{[1]}=\frac{1}{\kappa^{[0]}}[K^{[0]}-E_{\hat{k}}^{[0]2}],\nonumber\\
\nonumber\\
S_a^{\,\,b[3]}&=&\frac{1}{8\kappa^{[0]2}}[2\kappa^{[2]:b}_{:a}+\kappa^{[2]:a}_{:a}\delta_a^{\,\,b}-\kappa^{[0]}S^{[1]2}\delta_a^{\,\,b}]\nonumber\\
&+&\frac{1}{16\kappa^{[0]3}}[2E_{\hat{k}:a}^{[0]}E_{\hat{k}}^{[0]:b}-3E_{\hat{k}:c}^{[0]}E_{\hat{k}}^{[0]:c}\delta_a^{\,\,b}],\nonumber\\
\nonumber\\
S^{[3]}&=&\frac{1}{4\kappa^{[0]2}}[2\kappa^{[2]:a}_{:a}-\kappa^{[0]}S^{[1]2}]-\frac{1}{4\kappa^{[0]3}}E_{\hat{k}:a}^{[0]}E_{\hat{k}}^{[0]:a},\nonumber\\
\nonumber\\
E_{\hat{k}}^{[3]}&=&-\frac{1}{4\kappa^{[0]2}}[2\kappa^{[0]}S^{[1]}E_{\hat{k}}^{[0]}+E_{\hat{k}:a}^{[0]:a}].
\ea
Finally, we derive the first order expansions of the spacetime invariants near the horizon:
\ba
F^2&\approx&-\left.2E_{\hat{k}}^{2}\right\vert_{\hnH}+\left.\frac{1}{2\kappa^{2}}\hn\left[4K_{e}E_{\hat{k}}^{2}+E_{\hat{k}:a}^{2\,\,:a}-3E_{\hat{k}:a}E_{\hat{k}}^{\,\,:a}\right]\hspace{-0.07cm}\right\vert_{\hnH}\hn\hn k^2,\nonumber\\
\n{1.17a}\\
{\cal C}^2&\approx&\left.12K_e^2\right\vert_H-\frac{1}{\kappa^{2}}\hn\left[6K_e^2[3K_e-2E_{\hat{k}}^{2}]-[2K_e-E_{\hat{k}}^2]_{:a}\right.\nonumber\\
&\times&\left.\left.[2K_e-E_{\hat{k}}^2]^{:a}+6K_e[K_{e:a}^{\,\,\,\,:a}-2E_{\hat{k}}E_{\hat{k}:a}^{\,\,\,\,:a}]\right]\hspace{-0.07cm}\right\vert_{\hnH}\hn k^2,\nonumber\\
\n{1.17b}
\ea
where $\left.K_{e}\right\vert_{H}=[K-E_{\hat{k}}^2]\vert_{H}$.


\begin{thebibliography}{}
\bibitem{TD} Axisymmetric, distorted Schwarzschild black hole spacetime is of type I, and type D on the horizon and its axis of symmetry \cite{Pap}. It can be checked that the Harrison-Ernst transformation does not alter these properties, and in addition the corresponding spacetime is of type D on the Cauchy horizon.
\bibitem{f1} Discussion of the relation of these multipole moments to the  Newtonian and Geroch-Hansen relativistic multipole moments can be found in \cite{Ger1}-\cite{Que}.
\bibitem{f2} The metric \eq{2.1} allows Wick's rotation $T_{+}\to iT_E$ which transforms the exterior region $\eta\ge 1$ into that of the corresponding distorted Euclidean charged black hole.
The Euclidean horizon surface $\eta=1$ is a regular 2D totally geodesic surface. The electric field transforms as $\Phi_+\to -i\Phi_E$.
\bibitem{f3} It is known that an axisymmetric 2D surface can not be isometrically embedded into a 3D Euclidean space if its Gaussian curvature is negative at the vicinity of a fixed point of rotation \cite{Fr}. However, a failure of the embedding does not generally imply that Gaussian curvature of the surface is negative.
\end{thebibliography}
\end{document}